\documentclass[a4paper,fleqn,usenatbib]{mnras}

\usepackage{newtxtext,newtxmath}

\usepackage[T1]{fontenc}
\usepackage{ae,aecompl}

\usepackage{graphicx}
\usepackage{amsmath}
\usepackage{epsfig}
\usepackage{rotating}
\usepackage{color}
\usepackage{pdflscape}

\newcommand{\figdir}
  {figures/}
\newlength{\figwidth}
\newlength{\thinfigwidth}
\newlength{\midfigwidth}
\newcommand{\etc}
  {etc.}

\newcommand{\eg}
  {e.g.}
\newcommand{\cf}
  {cf.}
\newcommand{\ie}
  {i.e.}
\newcommand{\diff}
  {{\rmn{d}}}
\newcommand{\vect}[1]
  {\mbox{\boldmath ${#1}$}}

\newcommand{\eq}[1]
  {Eq.~(\ref{equation:#1})}

\newcommand{\sect}[1]
  {Section~\ref{section:#1}}

\newcommand{\tabl}[1]
  {{\mbox Table~\ref{table:#1}}}

\newcommand{\fig}[1]
  {Fig.~\ref{figure:#1}}

\newcommand{\prob}
  {{\rm{Pr}}}

\newcommand{\data}
  {\vect{d}}

\newcommand{\hubble}
  {H_0}

\newcommand{\step}
  {\Theta}
\newcommand{\metallicity}
  {[O/H]}
\newcommand{\uniform}
  {{\rm U}}
\newcommand{\normal}
  {{\rm N}}
\newcommand{\student}
  {{\rm T}}
\newcommand{\nuisance}
  {\vect{\theta}}

\defcitealias{Riess_etal:2016}{R16}
\newcommand{\riess}{\citetalias{Riess_etal:2016}}

\newcommand{\nul}{\nu_{\rm l}}
\newcommand{\nuc}{\nu_{\rm c}}
\newcommand{\tpeak}{t}
\newcommand{\studentdist}{student-t}
\newcommand{\psame}{\pi_\Lambda}
\newcommand{\pdiff}{\pi_{\bar{\Lambda}}}
\newcommand{\bayesfactor}{B}
\newcommand{\msame}{\Lambda}
\newcommand{\mdiff}{\bar{\Lambda}}

\newcommand{\hlobs}{\hat{h}_{\rm l}}
\newcommand{\hcobs}{\hat{h}_{\rm c}}
\newcommand{\hlerr}{\sigma_{\rm l}}
\newcommand{\hcerr}{\sigma_{\rm c}}
\newcommand{\datl}{d_{\rm l}}
\newcommand{\datc}{d_{\rm c}}

\newcommand{\slopep}{\ensuremath{s^p}}
\newcommand{\slopeZ}{\ensuremath{s^Z}}
\newcommand{\Mstdc}{\ensuremath{M^{\rm c}}}
\newcommand{\sigintc}{\ensuremath{\sigma^{\rm int,\,c}}}
\newcommand{\dmcal}{\ensuremath{\Delta m_i^{\rm cal}}}
\newcommand{\mc}{\ensuremath{m_{ij}^{\rm c}}}
\newcommand{\mhatc}{\ensuremath{\hat{m}_{ij}^{\rm c}}}
\newcommand{\phatc}{\ensuremath{\hat{p}_{ij}^{\rm c}}}
\newcommand{\Zhatc}{\ensuremath{\hat{Z}_{ij}^{\rm c}}}
\newcommand{\Mstds}{\ensuremath{M^{\rm s}}}
\newcommand{\sigints}{\ensuremath{\sigma^{\rm int,\,s}}}
\newcommand{\xsalt}{\ensuremath{x^{\rm s}}}
\newcommand{\csalt}{\ensuremath{c^{\rm s}}}
\newcommand{\xsalts}{\ensuremath{x_i^{\rm s}}}
\newcommand{\csalts}{\ensuremath{c_i^{\rm s}}}
\newcommand{\xsalthats}{\ensuremath{\hat{x}_i^{\rm s}}}
\newcommand{\csalthats}{\ensuremath{\hat{c}_i^{\rm s}}}
\newcommand{\alphas}{\ensuremath{\alpha^{\rm s}}}
\newcommand{\betas}{\ensuremath{\beta^{\rm s}}}
\newcommand{\msup}{\ensuremath{m_i^{\rm s}}}
\newcommand{\mhats}{\ensuremath{\hat{m}_i^{\rm s}}}
\newcommand{\dMLK}{\ensuremath{\Delta M_i^{\rm LK}}}
\newcommand{\decel}{\ensuremath{q_0}}
\newcommand{\zs}{\ensuremath{z_i^{\rm s}}}
\newcommand{\zhats}{\ensuremath{\hat{z}_i^{\rm s}}}
\newcommand{\dofc}{\ensuremath{\nu^{\rm c}}}
\newcommand{\dofs}{\ensuremath{\nu^{\rm s}}}
\newcommand{\tpeakc}{\ensuremath{\tpeak^{\rm c}}}
\newcommand{\tpeaks}{\ensuremath{\tpeak^{\rm s}}}
\newcommand{\kmsmpc}{\ensuremath{{\rm km\,s^{-1}\,Mpc^{-1}}}}
\newcommand{\dhubble}{\ensuremath{\Delta H_0}}
\newcommand{\ddecel}{\ensuremath{\Delta q_0}}
\newcommand{\hubblehat}{\ensuremath{\hat{H}_0}}
\newcommand{\decelhat}{\ensuremath{\hat{q}_0}}

\title[Hierarchical $\hubble$]
  {Clarifying the Hubble constant tension with a Bayesian 
  hierarchical model of the local distance ladder}

\author[S.\ M. Feeney et al.]{
Stephen M.\ Feeney$^{1,2}$\thanks{E-mail: sfeeney@simonsfoundation.org (SMF)},
Daniel J.\ Mortlock$^{2,3,4}$
and Niccol\`{o} Dalmasso$^{3,5}$
\\
$^1$Center for Computational Astrophysics, Flatiron Institute, 162 5th Avenue, New York, NY 10010, USA \\
$^2$Astrophysics Group, Imperial College London, Blackett Laboratory, Prince Consort Road, London SW7 2AZ, UK \\
$^3$Department of Mathematics, Imperial College London, London SW7 2AZ, UK \\
$^4$Department of Astronomy, Stockholm University, Albanova, SE-10691 Stockholm, Sweden \\
$^5$Department of Statistics, Carnegie Mellon University, Pittsburgh, PA 15213, USA
}

\date{Accepted 2017 ???? ??. 
  Received 2017 ???? ??; 
  in original form 2017 ???? ??}

\pubyear{2017}

\begin{document}
\label{firstpage}
\pagerange{\pageref{firstpage}--\pageref{lastpage}}
\maketitle

\begin{abstract}
Estimates of the Hubble constant, $\hubble$, from the local distance ladder and the cosmic microwave background (CMB) are discrepant at the $\sim$3-$\sigma$ level, indicating a potential issue with the standard $\Lambda$CDM cosmology. Interpreting this tension correctly requires a model comparison calculation which depends on not only the traditional `$n$-$\sigma$' mismatch but also the tails of the likelihoods. Determining the form of the tails of the local $\hubble$ likelihood is impossible with the standard Gaussian or least-squares approximation, as it requires using non-Gaussian distributions to faithfully represent anchor likelihoods and model outliers in the Cepheid and supernova (SN) populations, and simultaneous fitting of the complete distance-ladder dataset to ensure correct propagation of uncertainties. We have hence developed a Bayesian hierarchical model (BHM) that describes the full distance ladder, from nearby geometric-distance anchors through Cepheids to SNe in the Hubble flow. This model does not rely on any of the underlying distributions being Gaussian, allowing outliers to be modeled and obviating the need for any arbitrary data cuts. Sampling from the full $\sim$3000-parameter joint posterior distribution using Hamiltonian Monte Carlo and marginalizing over the nuisance parameters (\ie, everything bar $\hubble$), we find $\hubble$ = (72.72 $\pm$ 1.67) \kmsmpc\ when applied to the outlier-cleaned \cite{Riess_etal:2016} data, and ($73.15 \pm 1.78$) \kmsmpc\ with SN outliers reintroduced (the pre-cut Cepheid dataset is not available). Our high-fidelity sampling of the low-$\hubble$ tail of the distance-ladder posterior allows us to apply Bayesian model comparison to assess the evidence for deviation from $\Lambda$CDM. We set up this comparison to yield a lower limit on the odds of the underlying model being $\Lambda$CDM given the distance-ladder and \citet{Planck_XIII:2016} CMB data. The odds against $\Lambda$CDM are at worst 10:1 when considering the outlier-free distance-ladder data, or 7:1 when the SNe outliers are included and modeled, both considerably less dramatic than na\"ively implied by the $2.8$-$\sigma$ discrepancy. These odds become $\sim$60:1 when an approximation to the more-discrepant~\citet{Planck_Int_XLVI:2016} likelihood is included. The code used in this analysis is made publicly available at \url{https://github.com/sfeeney/hh0}.
\end{abstract}




\section{Introduction}
\label{section:intro}

The current expansion rate of the Universe is 
parametrized by the Hubble constant, 
denoted
$\hubble = 100 \, h \, \kmsmpc$,
with $h \simeq 0.7$.
As it
characterizes both the overall extra-Galactic distance scale
and the current age of the Universe,
the Hubble constant is one of the most important 
observable quantities in cosmology,
and so considerable resources have been devoted to measuring
$\hubble$ as precisely as possible
using a variety of independent methods.

One appealingly direct route to measuring $\hubble$ 
is to exploit local geometric distance measurements to calibrate 
brighter objects with known luminosity distances, working 
up through the distance ladder to objects with 
cosmological redshifts (see, e.g., \citealt{Freedman_etal:2001}, 
\citealt{Riess_etal:2009}, \citealt{Riess_etal:2011}, 
\citealt{Freedman_etal:2012} and \citealt{Riess_etal:2016}, 
and~\citealt{Freedman_Madore:2010} for a review).
By combining absolute distance measurements of NGC 4258, the 
Large Magellanic Cloud (LMC) and Milky Way (MW) Cepheids with 
observations of Cepheids in NGC 4258 and more distant
galaxies in which supernovae (SNe) have been identified, 
{\cite{Riess_etal:2016} (hereinafter \riess)
have determined that 
$\hubble = (73.24 \pm 1.74)$ \kmsmpc.
This 2.4\% relative uncertainty uncertainty includes
both the random `statistical'
contribution from measurement errors and the finite sample sizes 
along with the systematic
contribution that arises from, \eg, imperfect knowledge about the 
Cepheid period-luminosity-metallicity (P-L-Z) relationship.
The most important aspect of 
this (and any other) local $\hubble$ measurement 
is that all the sources involved are within $\sim 1\,{\rm Gpc}$, 
and so constitute 
a measurement of the current expansion rate of the Universe
that is almost completely independent of the cosmological expansion history.

It is also possible to measure -- or at least extrapolate -- 
a value for $\hubble$ from cosmological (i.e., higher redshift) 
measurements if certain assumptions are made about the 
cosmological model. 
The most precise cosmological estimates of $\hubble$
at present are based on
cosmic microwave background 
(CMB)
observations, which measure an angular
scale for the peaks in the power spectrum that are associated with 
known physical scales at recombination.
This can then be linked to the current expansion rate
by either assuming or inferring the 
cosmological expansion history in the time since recombination.
Adopting the standard spatially-flat `$\Lambda$CDM' model,
in which the cosmological expansion is dominated by a cosmological constant 
and cold dark matter,
the \cite{Planck_Int_XLVI:2016}
used systematics-cleaned large-scale polarization CMB measurements to obtain
$\hubble = (66.93 \pm 0.62)$ \kmsmpc.

That the locally-measured and CMB-derived estimates of $\hubble$
agree to within $\sim 7\%$ is a remarkable success
for modern observational cosmology, but the quoted uncertainties 
are considerably smaller than this difference, implying 
discrepancy at the 3.4-$\sigma$ level. 
This disagreement could indicate 
new physics beyond $\Lambda$CDM
(see, \eg,~\citealt{Planck_XIII:2016} for a discussion),
but of course there remains the possibility that the explanation
is more prosaic, and that there is instead a problem with 
the assumptions leading to one (or both!) of these two estimates. 
Reanalyses of the cosmological data have focused on {\it Planck's} 
consistency, both internally and with external datasets 
\citep{Spergel_etal:2015,Addison_etal:2016}, identifying splits 
in the data that produce conflicting cosmological conclusions 
(though see also \citealt{Planck_Int_LI:2016}).
One aspect of the local estimates that has come under scrutiny
is the treatment of the Cepheid data: 
\cite{Efstathiou:2014} argued that an improved approach to 
outlier rejection applied to the \cite{Riess_etal:2011}
data produced a lower estimate of $\hubble = (72.5 \pm 2.5) \, \kmsmpc$,
reducing the disagreement to 1.9-$\sigma$;
and
\cite{Cardona_etal:2016} used a heavier-tailed distribution to model
the intrinsic width of the P-L-Z relation to obviate the need for outlier
removal, leading to results that are consistent with those obtained by \riess.
More recently, \citet{Zhang_etal:2017}
introduced a formalism for blind inference of $\hubble$, 
minimizing human-induced bias in the analysis at the cost of a $\sim$30\% 
increase in the uncertainty on $\hubble$.
The estimate of $\hubble$ obtained by applying
this method to the \cite{Riess_etal:2011} data 
remained significantly higher than the {\em Planck} value,
implying that the discrepancy is not a product of human biases.
Varying the treatment of Cepheid extinction and 
colours~\citep{Follin_Knox:2017} and replacing the optical SNe 
photometry with near-infrared observations~\citep{Dhawan_etal:2017} 
similarly yield $\hubble$ estimates consistent with \riess. 
\citet{Rigault_etal:2015} found that the dependence of 
SN luminosity on local star-formation rate could bias $\hubble$
high by almost 3 \kmsmpc, thus apparently explaining much
of the tension; 
however, a similar
analysis with a larger SN sample, selected using the same cuts 
employed in cosmological analyses, found no significant 
effect~\citep{Jones_etal:2015}.
Cosmic variance has also been investigated as a potential 
explanation for the different $\hubble$ estimates, and 
while this does have an effect it is too small to provide an
explanation \citep{Wu_Huterer:2017}.

Another possible resolution is more purely statistical in nature, 
as the figures given above -- particularly the measure of discrepancy -- 
implicitly assume Gaussian statistics for the two $\hubble$ estimates,
leaving interpretation of the derived discrepancy measure open to error.
Another significant statistical limitation of the \riess\ approach
is that the Cepheid and SN data are processed separately, 
whereas the structure of the distance ladder means that 
there is potential for significant shrinkage (\ie, reduction in 
uncertainties) if a global approach is adopted (see also~\citet{Zhang_etal:2017}).
These issues are addressed in this paper, 
starting with a purely statistical demonstration that the significance 
of the apparent discrepancy is driven by the 
strength of the tails of the (posterior)
distributions of the two $\hubble$ measurements (\sect{stats}),
not just the separation between the peaks, 
and hence that the correct assessment of any discrepancy requires 
knowledge of these full distributions.
The rest of the paper describes a fully Bayesian formalism
for inferring the local value of $\hubble$,
which requires the development of a generative 
hierarchical model for the 
MASER, Cepheid and SN data that can incorporate outliers (\sect{ladder}).
The existing analysis methods are cast as special cases of our
formalism (\sect{existing_algorithms}), 
but a sophisticated sampling approach is necessary to utilize the full model (\sect{model}). The~\riess\ 
sample (\sect{data}) is used as the basis for simulations (\sect{sim}) and then analyzed in full (\sect{results}).  Our conclusions and possible extensions to this work are then summarized (\sect{conc}).


\section{Hubble constant discrepancies}
\label{section:stats}

The most important question to be addressed here is statistical in nature:
how probable is it that
the $\Lambda$CDM model can be rejected given that 
the Hubble constant inferred from cosmological datasets by 
assuming $\Lambda$CDM
differs from the locally measured expansion rate?
The data on the local expansion rate, $\datl$, can be summarized by 
the estimated value
$\hlobs = 0.7324$ and its associated uncertainty of $\hlerr = 0.0174$
from \riess;
the cosmological data, $\datc$, can be summarized by the implied 
estimate 
$\hcobs = 0.6693$ and its associated uncertainty of $\hcerr = 0.0062$
from the \cite{Planck_Int_XLVI:2016}.\footnote{For brevity and clarity, 
in this section we use the dimensionless quantity $h$ rather than 
$\hubble$. When discussing simulations and results, we will revert 
to $\hubble$.} The fact that 
\begin{equation}
\frac{
\left|\hlobs - \hcobs\right| 
}
{
\left( \hlerr^2 + \hcerr^2 \right)^{1/2}
}
\simeq
3.4
,
\end{equation}
leads to the statement that 
there is a 3.4-$\sigma$ discrepancy.
But, reported in this way, the
quantification of the putative discrepancy is heuristic in nature;
what is really needed is statement of how probable it is,
given the available data and other assumptions, 
that these two values are inconsistent.

A tempting -- and commonly used -- option would be to 
apply a classical hypothesis test, 
adopting the null hypothesis $\msame$, that 
$\Lambda$CDM holds and so the standard analysis 
of a cosmological dataset would yield a valid estimate of $h$.
The (two-tailed) $p$-value corresponding to $3.4 \, \sigma$
is an apparently decisive $0.00067$, 
well below the $5\%$ threshold often 
used to denote a `significant' result,
although obviously less than the more stringent
5-$\sigma$ level that is the standard for a 
detection in, \eg, particle physics.
However, this value is potentially misleading for two
distinct reasons:
it rests on the assumption that the tails of the two sampling 
distributions (\ie, the likelihoods) are Gaussian;
and, more fundamentally, interpretation of the
$p$-value
is made difficult by the fact it is calculated under
the assumption that the null hypothesis is true.
The quantitative effect of this is to penalize the 
null hypothesis too strongly in most cases.
This is straightforward to show theoretically if the 
sampling distribution is (close to being) Gaussian, 
(\eg, \citealt{Berger_Delampady:1987}) 
and has also been observed empirically 
in numerous cases \citep{Johnson:2013}.

Both of these issues can be addressed by 
using the formalism of Bayesian model comparison to assess
the relative support for two distinct models:
$\msame$, in 
which $\Lambda$CDM holds, and so analysing a CMB
dataset in the standard way
would yield an estimate of $\hubble$ that is consistent with the local value;
and 
$\mdiff$, in which some other process influences 
the expansion history
and so the standard
analysis of CMB data would yield an 
estimate of $\hubble$ that does not correspond to the current
expansion rate.
Bayes's theorem, combined with the law of total probability,
implies that the (posterior) probabilities of the two models are
\begin{align}
\prob(\mdiff | \datl , \datc)
 & = \left[1 + \frac{1 - \pdiff}{\pdiff} \,
   \frac{\prob(\datl, \datc | \msame)}{\prob(\datl, \datc | \mdiff)}
    \right]^{-1}, \label{equation:pmd}\\
\prob(\msame | \datl , \datc) & = 1 - \prob(\mdiff | \datl , \datc),\label{equation:pms}
\end{align}
where $\pdiff$ is the prior probability ascribed to the $\mdiff$ model,
$\datl$ and $\datc$ are the local and cosmological datasets,
and 
$\prob(\datl, \datc | \msame)$ 
and
$\prob(\datl, \datc | \mdiff)$
are the (marginal\footnote{The
marginal likelihood is
also referred to as the model-averaged likelihood or,
particular in astronomy, as the (Bayesian) evidence.}) likelihoods
under the two models.
Both models have unspecified parameters:
they share $h$, the true value of Hubble's constant 
that gives the current expansion rate of the Universe;
but $\mdiff$ has a second parameter,
here denoted as $\Delta$, 
which is defined as how much the estimated value of 
Hubble's constant inferred by assuming $\Lambda$CDM 
would differ from the true value.
So perfect cosmological data would yield
$\hcobs = h + \Delta$, 
and so comparing the local measurements with the 
Planck data na\"ively implies that $\Delta \simeq -0.06$.

The marginal likelihood of each model 
is obtained by integrating over its parameters:
\begin{equation}
\label{equation:evsame}
\prob(\datl, \datc | \msame)
  = \int_0^\infty \diff h
   \, \prob(h | \msame) \, \prob(\datl | h) \, \prob(\datc | h),
\end{equation}
where $\prob(h | \msame)$ is the prior distribution for $h$;
and 
\[
\prob(\datl, \datc | \mdiff)
\]
\vspace*{-4mm}
\begin{eqnarray}
\label{equation:evdiff}
  & = & 
  \int_0^\infty \diff h \int_{-\infty}^\infty \diff \Delta \,
    \prob(h, \Delta | \mdiff) \, \prob(\datl | h)
    \, \prob(\datc | h + \Delta)
  \\
  & = & 
  \int_0^\infty \diff h \,
    \prob(h | \mdiff) \, \prob(\datl | h) 
    \int_{-\infty}^\infty \diff \Delta \,
    \prob(\Delta | \mdiff) \, \prob(\datc | h + \Delta),
  \nonumber
\end{eqnarray}
where the second expression follows from assuming that the 
prior on $h$ and $\Delta$ can be factorized as
$\prob(h , \Delta | \mdiff) = \prob(h | \msame) \, \prob(\Delta | \mdiff)$.
Assuming that the prior on $h$ is the same for the two models 
is reasonable scientifically and convenient mathematically as 
it means that $\msame$ and $\mdiff$ are nested models, with 
$\msame$ being reproduced within $\mdiff$ by setting $\Delta = 0$.
As $\mdiff$ has more flexibility 
it inevitably provides a better optimal fit to the 
data than $\msame$; but 
this also means it is less predictive than 
the simpler $\msame$ model.
The trade-off between these two effects is captured by 
the Bayes factor, 
$\bayesfactor = \prob(\datl, \datc | \msame) / \prob(\datl, \datc | \mdiff)$,
although the main focus here will be on the posterior probability
of $\msame$ as defined in \eq{pms}.

Evaluating $\prob(\msame | \datl , \datc)$ 
is dependent on several distinct inputs:
a value for $\pdiff$;
forms for the parameter prior distributions,
$\prob(h | \msame)$ and
$\prob(\Delta | \mdiff)$;
and forms for the likelihoods 
$\prob(\datl | h)$ and $\prob(\datc | h + \Delta)$.
The main focus of this paper is the actual form of these 
likelihoods as implied by the available measurements,
so generic, but plausible, assumptions are made for the 
other inputs.
In this case adopting different values of $\pdiff$ could 
be used to state a degree of faith in the standard cosmological 
model, with the required 
level of evidence in favour of model $\mdiff$ being 
raised or lowered accordingly.

There is no compelling form for the parameter prior distributions,
but whatever forms are adopted should encode what is known
about these quantities while being sufficiently generic that 
minimal extra information is injected into the problem.
As $h$ is common to both models,
the values adopted for its prior centre and width 
have only a minor effect on the calculation, provided that
$\prob(h | \msame)$ is reasonably constant over the range
$0.65 \la h \la 0.75$ demanded by the data.
That can be achieved here by adopting priors of the form
\begin{equation}
\prob(h | \msame) = \normal(h; h_\pi, \sigma_h^2),
\end{equation}
where $\normal(x; \mu, \sigma^2)$
is the standard normal
distribution (Eq.~\ref{equation:normal}) in $x$ with mean $\mu$ and
variance $\sigma^2$.
The default option is to take $h_\pi \simeq 0.7$ and $\sigma_h \simeq 0.06$,
where the prior sensitivity can still be checked by adopting other 
values for these two parameters. Conversely, $\Delta$ is allowed to be non-zero
only in $\mdiff$, and its prior should therefore be chosen with a little more care. In the 
limit of a completely uninformative (\ie, broad) prior on $\Delta$, $\mdiff$ is 
unpredictive, and its posterior probability will tend to zero, independent of the data. If the 
prior width is extremely small (and the prior is centered on $\Delta = 0$), $\mdiff$ and $\msame$ become 
indistinguishable. If, instead, the prior is chosen such that the observed $\Delta$ 
is likely, $\mdiff$ will be favoured over $\msame$. This suggests casting 
$\mdiff$ as a designer model: selecting a prior on $\Delta$ of
\begin{equation}
\prob(\Delta | \mdiff) = \normal(\Delta; 0, \sigma_\Delta^2),
\end{equation}
with $\sigma_\Delta = 0.06 \simeq |\hcobs - \hlobs|$ to maximize the 
probability of the alternative model (assuming we have no preference 
for the sign of the difference between local and cosmological expansion 
rates). Following this logic, we set $\psame=0.5$ 
(and hence $\pdiff = 0.5$ as well), the smallest value 
this should take, given $\msame$ is the {\it de facto} null model. 
That leaves the main focus on the likelihoods, 
with results given for both the generic/simple Gaussian assumption
(\sect{gausslik})
and heavier-tailed distributions 
(\sect{nongausslik}).


\subsection{Gaussian likelihoods}
\label{section:gausslik}

The default interpretation
of the two primary Hubble constant measurements is 
that the
likelihoods are Gaussian, so that 
$\prob(\datl | h) = \normal(\hlobs; h, \hlerr^2)$
and 
$\prob(\datc | h) = \normal(\hcobs; h + \Delta, \hcerr^2)$,
where $\hlobs$ and $\hlerr$ are implicitly dependent on $\datl$
and $\hcobs$ and $\hcerr$ are, similarly, dependent on $\datc$.
These Gaussian local and cosmological likelihoods are plotted in 
\fig{toy_model_likelihoods} as solid purple and dot-dashed gray lines, 
respectively. Inserting these expressions into \eq{evsame}
yields 
the marginal likelihood for the simpler model as
\begin{align}
\prob(\datl, \datc | \msame) = & \int_0^\infty \diff h \, \normal(h; h_\pi, \sigma_h^2) \, \normal(\hlobs ; h, \hlerr^2) \, \normal(\hcobs ; h, \hcerr^2) \nonumber \\
= &
\frac{1}{2\pi} \, \frac{1}{\sqrt{\sigma_h^2 \, \hlerr^2 + \sigma_h^2 \, \hcerr^2 + \hlerr^2\, \hcerr^2}} \nonumber \times \\
  &  \exp \left\{ -\frac{1}{2} \left[ \frac{ \left( h_\pi - \hlobs \right)^2 }{ \sigma_h^2 + \hlerr^2 + \frac{ \sigma_h^2 \, \hlerr^2 }{ \hcerr^2 } } +  \frac{ \left( h_\pi - \hcobs \right)^2 }{ \sigma_h^2 + \hcerr^2 + \frac{ \sigma_h^2 \, \hcerr^2 }{ \hlerr^2 } } + \right. \right. \nonumber \\
  & \phantom{\exp \{ -\frac{1}{2} [{}} \left. \left. \frac{ \left( \hlobs - \hcobs \right)^2 }{ \hlerr^2 + \hcerr^2 + \frac{ \hlerr^2 \, \hcerr^2 }{ \sigma_h^2 } } \right] \right\}.
\end{align}
For the more complicated model the marginal likelihood is,
from \eq{evdiff},
\begin{align}
\prob(\datl, \datc | \mdiff) = & \int_0^\infty \diff h \, \normal(h; h_\pi, \sigma_h^2) \,  \normal(\hlobs ; h, \hlerr^2) \times  \nonumber \\
  & \int_{-\infty}^\infty \diff \Delta \, \normal(\Delta; 0, \sigma_\Delta^2) \, \normal(\hcobs ; h + \Delta, \hcerr^2) \nonumber \\
= &
\frac{1}{2\pi} \, \frac{1}{\sqrt{\sigma_h^2 \, \hlerr^2 + \sigma_h^2 \, \left(\hcerr^2 + \sigma_\Delta^2 \right) + \hlerr^2\, \left(\hcerr^2 + \sigma_\Delta^2 \right)}} \nonumber \times \\
  &  \exp \left\{ -\frac{1}{2} \left[ \frac{ \left( h_\pi - \hlobs \right)^2 }{ \sigma_h^2 + \hlerr^2 + \frac{ \sigma_h^2 \, \hlerr^2 }{ \hcerr^2 + \sigma_\Delta^2 } } + \right. \right. \nonumber \\
  &  \phantom{\exp \{ -\frac{1}{2} [{}} \frac{ \left( h_\pi - \hcobs \right)^2 }{ \sigma_h^2 + \hcerr^2 + \sigma_\Delta^2 + \frac{ \sigma_h^2 \, \left(\hcerr^2 + \sigma_\Delta^2 \right) }{ \hlerr^2 } } + \nonumber \\
  & \phantom{\exp \{ -\frac{1}{2} [{}} \left. \left. \frac{ \left( \hlobs - \hcobs \right)^2 }{ \hlerr^2 + \hcerr^2 + \sigma_\Delta^2 + \frac{ \hlerr^2 \, \left(\hcerr^2 + \sigma_\Delta^2 \right) }{ \sigma_h^2 } } \right] \right\}.
\end{align}
These marginal likelihoods combine to give the Bayes factor 
\begin{align}
B = & \sqrt{\frac{\hcerr^2 + \sigma_\Delta^2}{\hcerr^2} \, \frac{\sigma_h^{-2} + \hlerr^{-2} + \left(\hcerr^2 + \sigma_\Delta^2 \right)^{-1} }{ \sigma_h^{-2} + \hlerr^{-2} + \hcerr^{-2}}} \nonumber \times \\
  &  \exp \left\{ -\frac{1}{2} \left[ \frac{ - \sigma_\Delta^2 \, \sigma_h^2 \, \hlerr^2 \, \left( h_\pi - \hlobs \right)^2 }{  \hcerr^2 \left( \hcerr^2 + \sigma_\Delta^2 \right) \left(\sigma_h^2 + \hlerr^2 + \frac{ \sigma_h^2 \, \hlerr^2 }{ \hcerr^2 } \right) \left(\sigma_h^2 + \hlerr^2 + \frac{ \sigma_h^2 \, \hlerr^2 }{ \hcerr^2 + \sigma_\Delta^2 } \right) } + \right. \right. \nonumber \\
  &  \phantom{ \exp ( -\frac{1}{2} [{}} \frac{ \sigma_\Delta^2 \left(1 + \frac{ \sigma_h^2 }{ \hlerr^2} \right) \left( h_\pi - \hcobs \right)^2 }{ \left( \sigma_h^2 + \hcerr^2 + \frac{ \sigma_h^2 \,\hcerr^2 }{ \hlerr^2 } \right) \left (\sigma_h^2 + \hcerr^2 + \sigma_\Delta^2 + \frac{ \sigma_h^2 \, \left(\hcerr^2 + \sigma_\Delta^2 \right) }{ \hlerr^2 } \right) } + \nonumber \\
  & \phantom{\exp ( -\frac{1}{2} [{}} \left. \left. \frac{ \sigma_\Delta^2 \left(1 + \frac{ \hlerr^2 }{ \sigma_h^2 } \right) \left( \hlobs - \hcobs \right)^2 }{ \left( \hlerr^2 + \hcerr^2 + \frac{ \hlerr^2 \,\hcerr^2 }{ \sigma_h^2 } \right) \left (\hlerr^2 + \hcerr^2 + \sigma_\Delta^2 + \frac{ \hlerr^2 \, \left(\hcerr^2 + \sigma_\Delta^2 \right) }{ \sigma_h^2 } \right) } \right] \right\}.
\end{align}

At first glance, these equations show that the marginal likelihoods and Bayes factor depend solely, but rather opaquely, on the prior widths and tensions between the values of $h$ preferred by the prior and two likelihoods. In the asymptotic regime that $\sigma_h$ is much larger than $\hlerr$, $\hcerr$ and the differences between means, however, these forms simplify greatly. The Bayes factor in particular becomes independent of the prior on $h$, depending only on the tension between the two measurements, their uncertainties, and the width of the prior on $\Delta$ as
\begin{equation}
B \simeq  \sqrt{\frac{\hlerr^2 + \hcerr^2 + \sigma_\Delta^2}{\hlerr^2 + \hcerr^2}} \exp \left\{ -\frac{1}{2} \left[  \frac{ \sigma_\Delta^2 }{ \hlerr^2 + \hcerr^2 + \sigma_\Delta^2 } \, \frac{ \left( \hlobs - \hcobs \right)^2 }{ \hlerr^2 + \hcerr^2 } \right] \right\}.
\end{equation}

\begin{figure}
\includegraphics[width=\figwidth]{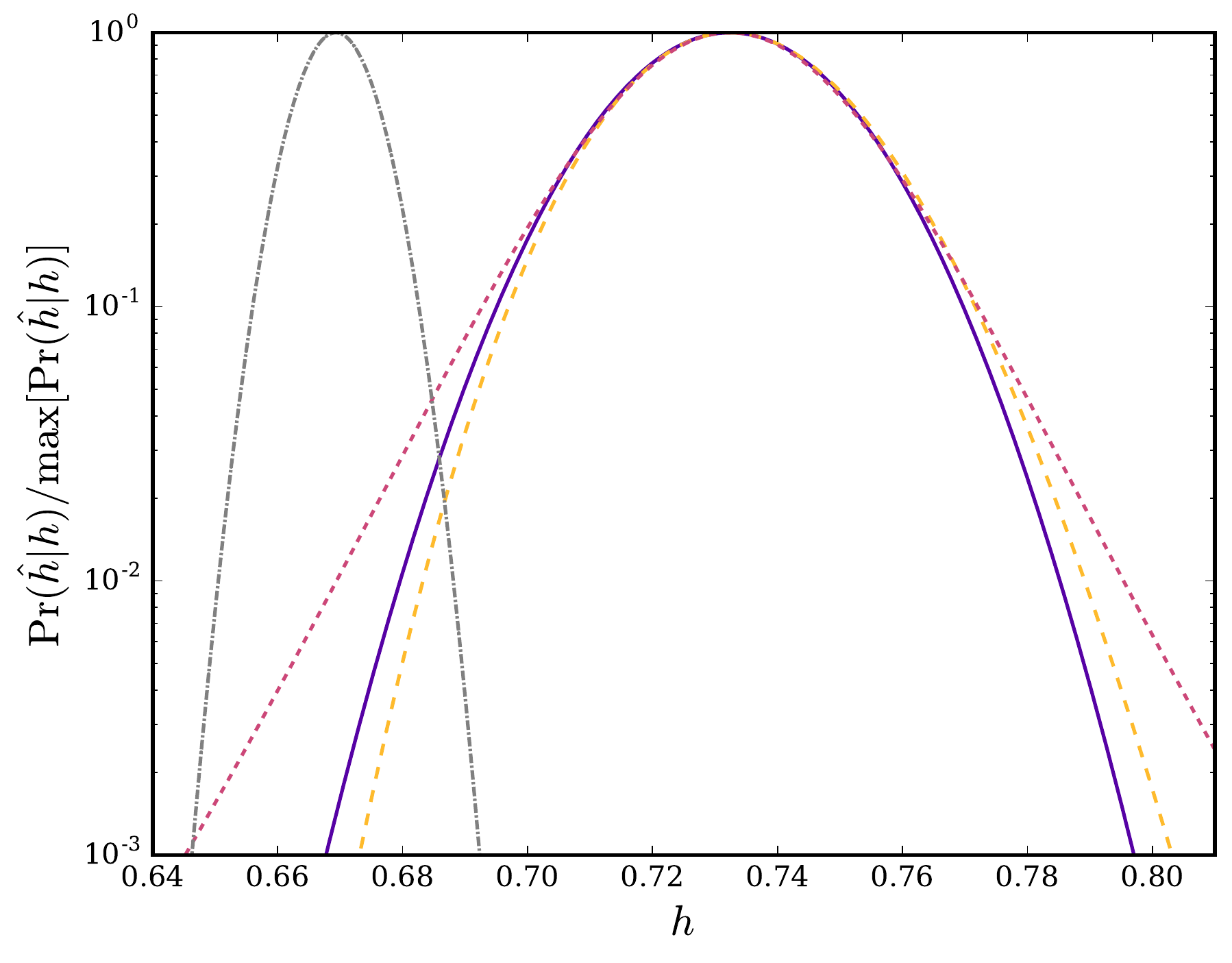}
\caption{Example likelihoods used in pedagogical model comparison calculations: Gaussian local $h$ measurement (purple solid), Gaussian MASER distance (yellow long-dash), student-t local $h$ measurement with $\nu=10$ (pink short-dash) and {\it Planck} CMB estimate (gray dot-dash).}
\label{figure:toy_model_likelihoods}
\end{figure}

\begin{figure}
\includegraphics[width=\figwidth]{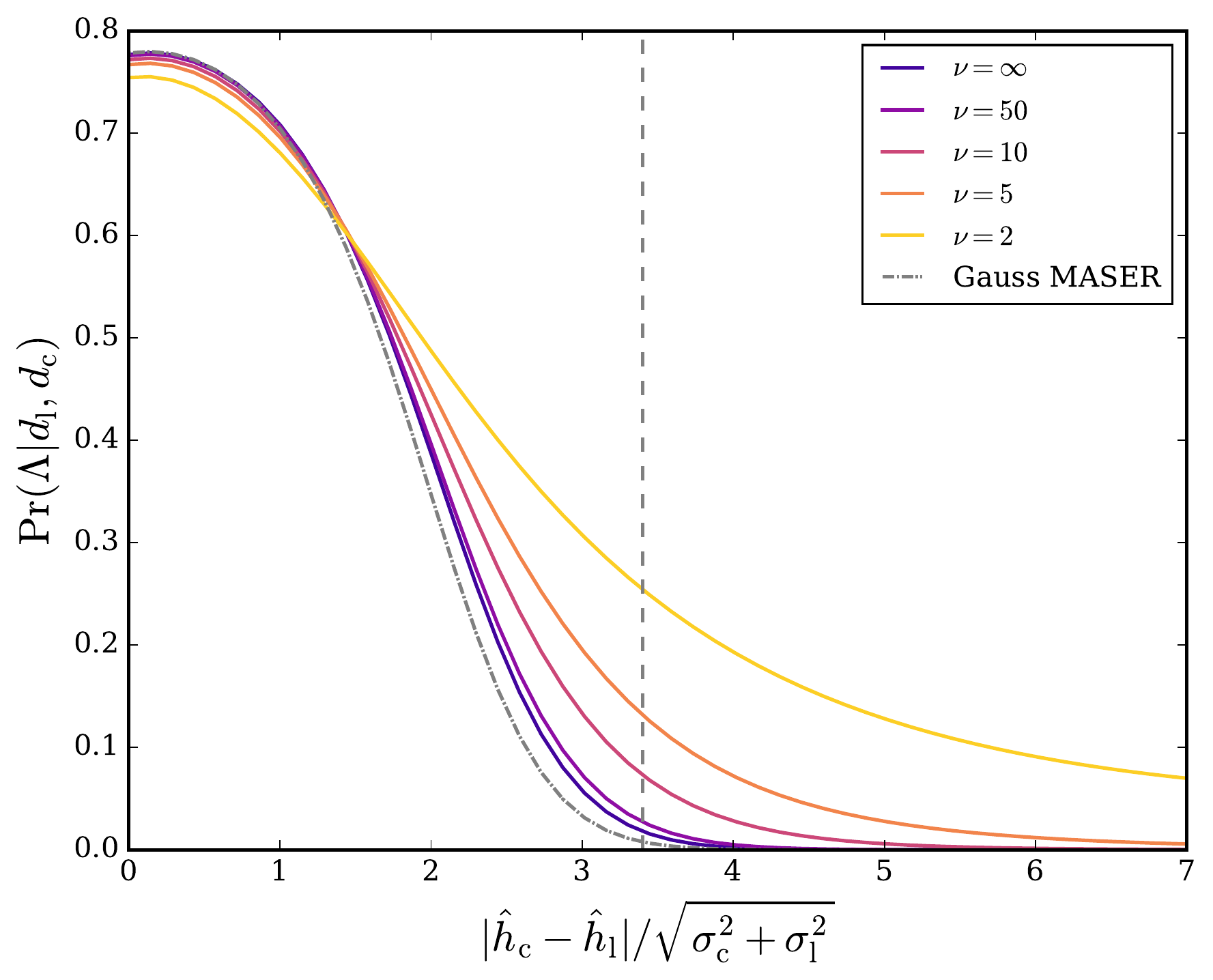}
\includegraphics[width=\figwidth]{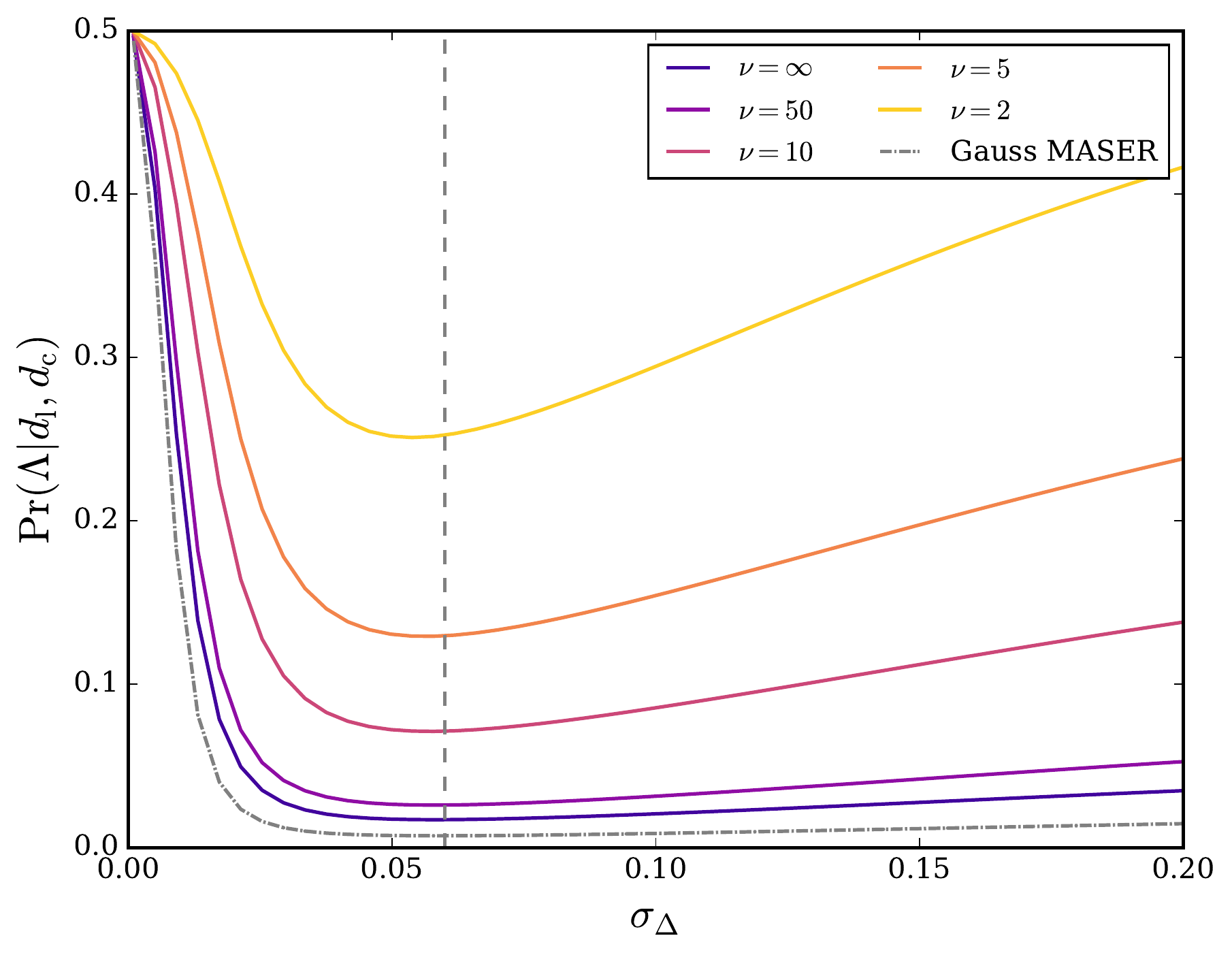}
\caption{Top: posterior probability of our standard model, $\Lambda$, versus `tension' for increasingly heavy-tailed likelihoods (coloured solid lines) and Gaussian MASER distance likelihood (gray dot-dashed line). Bottom: as above, but plotted as a function of the width of the prior on the difference between the local and cosmological $h$ values.}
\label{figure:toy_model_evidences}
\end{figure}

The dependence of $\prob(\Lambda | \datl, \datc)$ on the `tension', $|\hcobs - \hlobs| / (\hcerr^2 + \hlerr^2)^{1/2}$, is plotted in the top panel of \fig{toy_model_evidences} in dark blue,
assuming the prior choices detailed above. The tension is varied by adjusting the mean of the cosmological likelihood, keeping all other parameters fixed. Picking out the observed tension between \riess\ and the \citet{Planck_Int_XLVI:2016}, 3.4, we see this corresponds to a Bayes factor of 0.017. The dependence of the model probability on the width of the $\Delta$ prior is plotted in the bottom panel of \fig{toy_model_evidences}, with the minimum in probability clearly located at prior widths roughly equal to the observed discrepancy, corresponding to the $\mdiff$ model designed to do just this. We would therefore state that the minimum posterior probability of $\msame$ is 0.017 given these data. If instead we compare \riess\ with the $h$ estimated by the \citet{Planck_XIII:2016} -- a tension of 2.8 -- we find a Bayes factor of 0.12, or a minimum model probability of 0.10. Though these model probabilities strongly disfavour the standard model (by odds ratios of 1:60 and 1:10, respectively), they are much less extreme than the corresponding $p$-values would suggest.
Additionally, these probabilities have a clear and unambiguous interpretation.


\subsection{Non-Gaussian likelihoods}
\label{section:nongausslik}

There is no fundamental reason why the $h$-dependence
of either the local or cosmological likelihood should be precisely Gaussian.
Deviations from Gaussianity
in $\prob(\datl | h )$ 
for $h \simeq \hcobs$
or in 
$\prob(\datc | h + \Delta)$ 
for $h + \Delta \simeq \hlobs$
would affect the values of the marginal likelihoods and hence the 
posterior probabilities of the two models.
To illustrate this we consider two plausible examples of non-Gaussian 
likelihoods:
that resulting from a Gaussian uncertainty in the distance to a 
single anchor
(\sect{maser_gauss});
and 
a generic heavy-tailed distribution 
(\sect{heavy_tailed}).


\subsubsection{Gaussian MASER distance}
\label{section:maser_gauss}

A concrete motivating example for a non-Gaussian form of the 
local Hubble constant constraints would come about if the 
distance ladder had just a single rung, in the slightly artificial
form of MASER with a perfectly known cosmological redshift.
It is reasonable to assume that the measurements of the MASER 
result in a likelihood of the form
$\prob(\datl | D_{\rm M})
  = \normal(\hat{D}_{\rm M}; D, \sigma_{\rm M}^2)$,
where the MASER distance estimate $\hat{D}_{\rm M}$
and the uncertainty $\sigma_{\rm M}^2$ are both implicitly functions 
of the local data, $\datl$.
Given the known (cosmological) redshift, $z_{\rm M}$, 
the natural estimate of $h$ would be 
$\hat{h}_{\rm M} = c \, z_{\rm M} / (H_{100} \, \hat{D}_{\rm M})$,
with the associated uncertainty 
$\hat{\sigma}_{h} = \sigma_{\rm M} / (c \, z_{\rm M} / H_{100})$,
where $H_{100} = 100 \, \kmsmpc$.
But the MASER likelihood, treated as a function of $h$, 
would not be Gaussian, instead having the form
\begin{align}
\prob(\datl | h) & = 
  \frac{1}{(2 \pi)^{1/2} \, \sigma_{\rm M}}
  \exp\left[-\frac{1}{2} 
  \frac{\left( 1/\hat{h} - 1/h\right)^2}{\hat{\sigma}_h^{2}}
  \right] \\
  & \propto \normal(1/\hat{h}_{\rm M}; 1/h, \hat{\sigma}_h^2).\nonumber
\end{align}
This likelihood is plotted as a yellow dashed line in \fig{toy_model_likelihoods}, choosing $\hat{\sigma}_h$ such that the resultant $h$ distribution's variance matches $\hlerr^2$. The resulting skew towards higher values of $h$ results in an increased discrepancy between the local and cosmological measurements, despite the fact that $\hlobs$ and $\hlerr$ are unchanged. This effect is highlighted  in \fig{toy_model_evidences}, in which the posterior probability of $\msame$ is lower for the Gaussian $D_{\rm M}$ likelihood than the Gaussian $\hlobs$ likelihood except when the local and cosmological estimates differ by less than 1-$\sigma$.


\subsubsection{Heavy-tailed distribution}
\label{section:heavy_tailed}

It is also instructive to consider a more generic heavy-tailed
distribution for the measurement of $h$.
One flexible option is a scaled and translated
student-t distribution, 
$\student(x; \mu, \sigma, \nu)$,
defined in \eq{student}.
The likelihoods for the two datasets are now
$\prob(\datl | h) = \student(\hlobs; h, \hlerr, \nul)$
and 
$\prob(\datc | h, \Delta) = \student(\hcobs; h + \Delta, \hcerr, \nuc)$,
where $\nul$ and $\nuc$ encode how much heavier the tails 
of the two likelihoods are relative to the idealized normal model
discussed in \sect{gausslik} (and recovered if $\nu \rightarrow \infty$).
An example likelihood with $\nu=10$ is plotted in 
\fig{toy_model_likelihoods} as a pink short-dashed line. The standard 
deviation of this likelihood is only 11\% larger than a Gaussian with 
the same scale, but it predicts extreme events should occur with 
significantly higher frequency. The marginal likelihoods must now 
be calculated numerically, but this is simple within this two-dimensional setup.

The results of adopting this model are shown in 
\fig{toy_model_evidences} for a range of $\nul=\nuc = \nu$ values,
demonstrating in particular the expected effect that the 
null hypothesis, model $\Lambda$, is much less strongly disfavoured
if the likelihoods are heavy-tailed. Taking the 
current tension between the local and cosmological $h$ estimates as an 
example, we see that $\msame$ is 15 times more probable if the two 
likelihoods are student-t with $\nu=2$ than if they are Gaussian. 
The tails of the local and cosmological $h$ likelihoods play a critical 
role in understanding the true tension between datasets. Assessing
this discrepancy therefore requires a method of calculating these full 
distributions, rather than assuming them to be Gaussian.


\section{A generative model of local distance ladder data}
\label{section:ladder}

Motivated by the above demonstration that any assessment of 
discrepancy between the local and cosmological estimates of $\hubble$
is dependent on knowing the full form of the likelihoods,
we set about developing a method to determine this for the 
local distance ladder.
There are a number of options for which rungs are used,
but here we focus on the 
combination of MASERs, Cepheids and SNe 
used by 
\cite{Riess_etal:2009,Riess_etal:2011,Riess_etal:2016} to estimate $\hubble$.
Before even considering how such a complicated
dataset might be analyzed to 
estimate $\hubble$, an important first step is to describe the 
links between the underlying parameters, observable quantities 
and actual measured data.  
In statistical terms this means developing a generative model
which, if it incorporated accurate
descriptions of all the relevant phenomena,
could be used to generate a realistic dataset that,
in particular, predicts outliers as well as the core distributions.
Such a model must hence describe both the relevant
astrophysical objects and the various measurement processes.

Such a model is inevitably complicated, and we start by listing all the relevant quantities\footnote{The notation used here differs from that adopted by \riess\ as additional parameters are required here.  Both our notation and that used by \riess\ are listed in \tabl{params}.} in \tabl{params}. The links between these quantities are illustrated, in the form of a directed acyclic graph, in \fig{network_full}.  The nodes of the graph consist of: the observed data (double circles); the model parameters describing individual objects (single circles); the astronomical populations (blue and red rectangles); and, ultimately, the cosmological parameters, including $\hubble$.  The links between these quantities are, in general, stochastic, and so described by probability distributions (orange rectanges) that describe both astrophysical phenomena (most obviously the intrinsic scatter of the Cepheid P-L-Z relationship) and measurement processes.

Visualizing the distance ladder in this way is enlightening for two distinct reasons. Reading the model `forwards' (i.e., starting at the top of \fig{network_full} and moving down) reveals the data generation process, starting with global parameters, moving through populations to individual objects and then finally to actual measured quantities.  This is the generative model for the data, which could be used to create simulations. Reading the model `backwards' (i.e., starting with the data at the bottom and moving upwards) demonstrates the structure of the data analysis task that allows the global parameters of interest to be inferred from the measurements of the anchors, Cepheids and SNe.  This provides the basis for the Bayesian formalism described in \sect{model}.

The distance ladder has three main components:
\begin{enumerate}
\item
Objects with geometrical distance measurements, or anchors (the left-most of the three blue rectangles in \fig{network_full}). Three types of anchors are employed in~\riess's preferred analysis: the absolute distance to NGC~4258 inferred from MASER
measurements by~\cite{Humphreys_etal:2013}; parallaxes of 15 Milky Way Cepheids~\citep{vanLeeuwen_etal:2007,Riess_etal:2014,Casertano_etal:2015}; and the absolute distance to the LMC, derived from detached eclipsing binary observations~\citep{Pietrzynski_etal:2013}.
\item
The periods, metallicities and (average) apparent Wesenheit magnitudes~\citep{Madore:1982} of a large number of Cepheids in galaxies which are either anchors or which host a Type Ia SN (the middle of the three blue rectangles in \fig{network_full}). Observing these Cepheid properties allows inference of their luminosity (and hence distance) via the Cepheid period-luminosity relationship~\citep{Leavitt:1908,Leavitt_Pickering:1912} or Leavitt Law, which has since been extended to (potentially) also include metallicity dependence~\citep{Freedman_Madore:2011}.

\item
Type Ia SNe in both the Cepheid host galaxies and more distant galaxies in the Hubble flow (the right-most of the three blue rectangles in \fig{network_full}). The final catalogue represents a set of SNe with cosmological redshifts and peak magnitudes calibrated to one or more absolute distance measures. Comparing the redshift measurements of these galaxies with their SN distances yields a tight bound on the Hubble constant.
\end{enumerate}

At the heart of the distance ladder lie a small number of equations. The distance, $d_i$, and parallax, $\pi_i$, of the $i^{\rm th}$ anchor host are related to its distance modulus, $\mu_i$, by
\begin{equation}
\mu_i= 5 \log_{10}{\left( \frac{d_i}{\rm pc} \right)} - 5 = -5 \log_{10}{\left( \frac{\pi_i}{\rm arcsec} \right)} - 5 - \dMLK \label{equation:anchor}
\end{equation}
where $\dMLK$ is a bias correction proposed by~\citet{Lutz_Kelker_1973} to account for the preferential scattering of distant stars into parallax surveys. The distance modulus relates a source's apparent and absolute magnitudes via $\mu = m - M$.

The observed correlation between the period, $p_{ij}$, metallicity, $Z_{ij}$, and apparent magnitude, $\mc$, of the $j^{\rm th}$ Cepheid in the $i^{\rm th}$ host can be written as
\begin{equation}
\mc = \mu_i + \Mstdc + \slopep \log_{10} (p_{ij}) + \slopeZ \log_{10} Z_{ij} \,(\,+\,\dmcal),
\label{equation:cepheid_pzl}
\end{equation}
where
$\Mstdc$ is the absolute magnitude of a Cepheid with a one-day period and solar metallicity, and $\dmcal$ is an artificial zero-point offset potentially introduced by observing Cepheids in different hosts with different instruments. The observed scatter around Equation~\ref{equation:cepheid_pzl} is typically assumed to be normally distributed around zero with standard deviation $\sigintc$, though the presence of outliers (\eg, crowded or anomalous Cepheids) may alter the precise form of this distribution. If sigma clipping is employed to remove outliers, this distribution should strictly be truncated. We do not include a break in the period slope~\citep{Ngeow_Kanbur:2005,Kodric_etal:2015} in this analysis, as \riess\ find no evidence for this; it is simple to add if required.

\begin{landscape}
\begin{figure}
\includegraphics[width=\columnwidth]{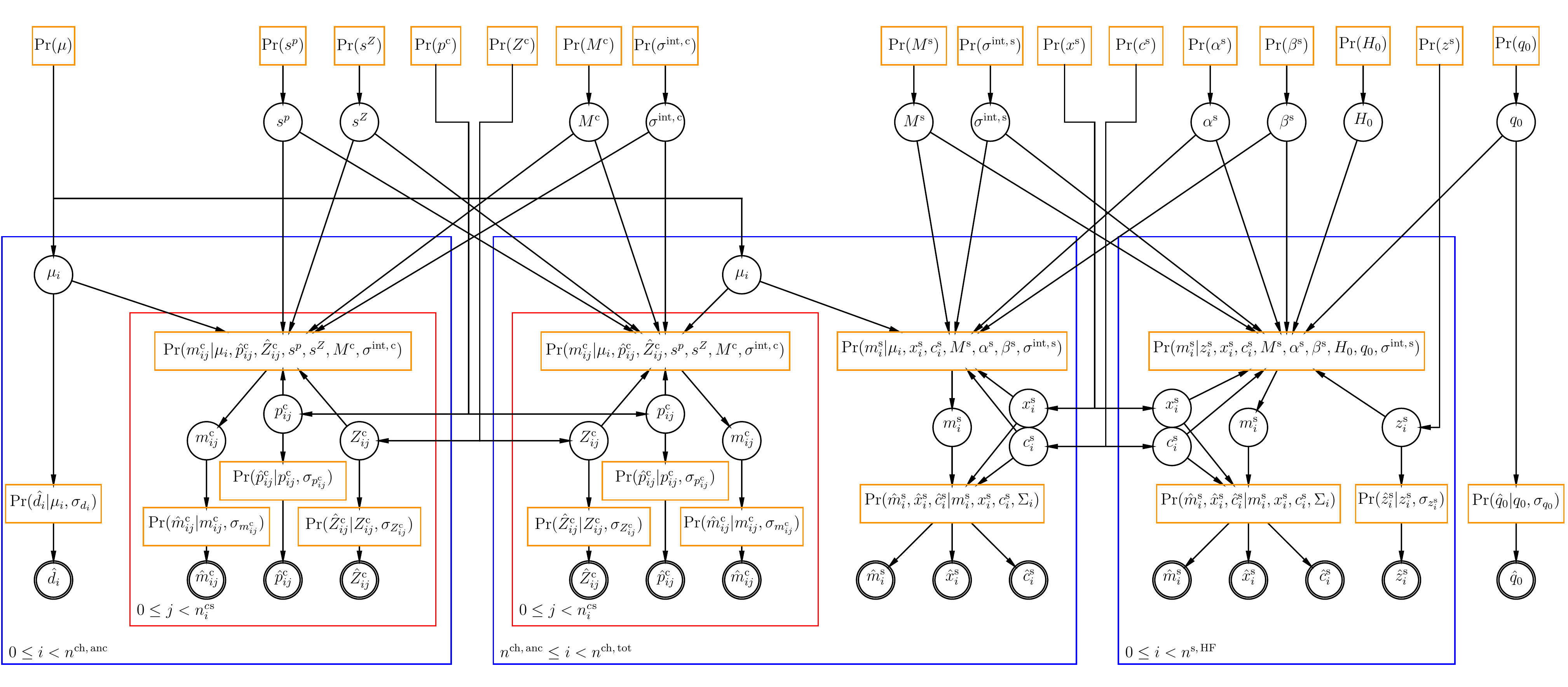}
\caption{The complete Bayesian network for the simultaneous analysis of the anchor, Cepheid and SN data within the model described by Equations~\ref{equation:anchor}-\ref{equation:z2mu}, shown as a directed acyclic graph. Data are plotted as double black circles, model parameters as single black circles. Probability distributions are indicated by orange rectangles, Cepheid populations with red boxes and host populations with blue boxes.}
\label{figure:network_full}
\end{figure}
\end{landscape}

The dependence of the brightness of a Type Ia SNe on its light-curve shape ($\xsalts$) and colour ($\csalts$) is captured\footnote{In its simplest form: see, \eg, \citet{Marriner_etal:2011}, \citet{Shariff_etal:2016} and \citet{Mandel_etal:2016} for extensions.} by the Tripp relation~\citep{Phillips:1993,Tripp:1998},
\begin{equation}
\msup = \mu_i + \Mstds + \alphas \xsalt_i + \betas \csalt_i,
\label{equation:tripp}
\end{equation}
where $\Mstds$ is the standard absolute magnitude of Type Ia SNe. As with the Cepheids, the intrinsic scatter around this relation is modeled as a normal distribution with standard deviation $\sigints$. For a Hubble flow SN with cosmological redshift $\zs$, we can finally relate the distance modulus to redshift, and hence make contact with cosmological quantities, using the expansion
\begin{align}
\mu_i = 5 \log_{10} \left\{ \frac{c\,\zs}{\hubble \, {\rm Mpc}} \right. & \bigg[ 1 + \frac{1}{2}(1 - \decel) \zs - \nonumber\\
& \left.\left. \frac{1}{6}\left(1 - \decel - 3\decel ^ 2 + j_0^2\right) (\zs)^2 \right] \right\} + 25,
\label{equation:z2mu}
\end{align}
where $\decel$ is the deceleration parameter, $j_0$ is the jerk (fixed to unity by the assumption of a flat $\Lambda$CDM cosmology), and $c$ is the speed of light.

The rungs of the distance ladder are climbed through Equations~\ref{equation:anchor}-\ref{equation:z2mu}: geometrical distance measurements, converted to distance moduli, calibrate the standard absolute magnitudes of nearby Cepheids and more-distant SNe, reducing the uncertainty on the Hubble Constant derived from SNe at cosmological distances. An important part of this scheme is the handling of uncertainties, which come from both stochastic astrophysical processes and measurement errors.  Our aim is to include all these sources of uncertainty in our final inference for $\hubble$, but before describing our scheme for doing this it is important to review the existing approaches for obtaining local $\hubble$ estimates.


\begin{table}
\centering
\scriptsize
\caption{Model parameters, data and constants.}
\label{table:params}
\begin{tabular}{lll}
\hline
quantity &
  \riess & 
      definition \\
\hline
$\mu_i$ &
  $\mu_{0,i}$ &
      distance modulus of $i^{\rm th}$ host \\
$\hat{d}_i$ &
  - &
      measured distance of anchor \\
$\sigma_{d_i}$ &
  - &
      uncertainty in anchor distance \\
$\hat{\pi}_i$ &
  - &
      measured parallax of anchor \\
$\sigma_{\pi_i}$ &
  - &
      uncertainty in anchor parallax \\
$\dMLK$ &
  - &
      bias in parallax anchor magnitude \\
\hline
$\phatc$ &
  $P_{ij}$ &
      measured period of $j^{\rm th}$ Cepheid in $i^{\rm th}$ host \\
$\Zhatc$ &
  $(O/H)_{ij}$ &
      measured metallicity (\metallicity) of $j^{\rm th}$ Cepheid \\
 & & in $i^{\rm th}$ host \\
$\mc$ &
  $m_{H,ij}^W$ &
      apparent magnitude of $j^{\rm th}$ Cepheid in $i^{\rm th}$ host \\
$\mhatc$ &
  - &
      measured apparent magnitude of $j^{\rm th}$ Cepheid \\
 & & in $i^{\rm th}$ host \\
$\sigma_{\mc}$ &
  - &
      uncertainty in apparent magnitude of $j^{\rm th}$ \\
 & & Cepheid in $i^{\rm th}$ host \\
$\slopep$ &
  $b_W$ &
      log-period slope of Cepheid Leavitt Law \\
$\slopeZ$ &
  $Z_W$ &
      log-metallicity slope of Cepheid Leavitt Law \\
$\Mstdc$ &
  $M_{H,1}^W$ &
      absolute magnitude of solar-metallicity \\
 & & Cepheid with one-day period \\
$\sigintc$ &
  $\sigma_{\rm int}$ &
      intrinsic scatter about Cepheid Leavitt Law \\
$\dofc$ &
  - &
      degrees of freedom in heavy-tailed scatter \\
 & & about Cepheid Leavitt Law \\
$\tpeakc$ &
  - &
      peak density of \studentdist\ distribution with $\dofc$ \\
 & & relative to Gaussian with same location/scale\\
$\dmcal$ &
  $\Delta$zp &
      offset between ground- and space-based \\
 & & magnitudes \\
\hline
$\msup$ &
  $m_{x,i}^0$ &
      peak apparent magnitude of SN in $i^{\rm th}$ host \\
$\mhats$ &
  - &
      estimated peak apparent magnitude of SN in \\
 & & $i^{\rm th}$ host \\
$\xsalts$ &
  - &
      light-curve stretch of SN in $i^{\rm th}$ host \\
$\xsalthats$ &
  - &
      estimated light-curve stretch of SN in $i^{\rm th}$ host \\
$\csalts$ &
  - &
      light-curve colour of SN in $i^{\rm th}$ host \\
$\csalthats$ &
  - &
      estimated light-curve colour of SN in $i^{\rm th}$ host \\
$\Sigma_i$ &
  - &
      covariance matrix of $i^{\rm th}$ SN's estimated light- \\
 & & curve parameters \\
$\zs$ &
  $z_i$ &
      redshift of $i^{\rm th}$ galaxy \\
$\zhats$ &
  - &
      measured redshift of $i^{\rm th}$ galaxy \\
$\sigma_{\zs}$ &
  - &
      uncertainty in measured redshift of $i^{\rm th}$ galaxy \\
$\alphas$ &
  - &
      light-curve stretch coefficient in SN Tripp \\
 & & relation \\
$\betas$ &
  - &
      light-curve colour coefficient in SN Tripp \\
 & & relation \\
$\Mstds$ &
  $M_x^0$ &
      absolute magnitude of Type Ia SN \\
$\sigints$ &
  - &
      intrinsic scatter about SN Tripp relation \\
$\dofs$ &
  - &
      degrees of freedom in heavy-tailed scatter \\
 & & about SN Tripp relation \\
$\tpeaks$ &
  - &
      peak density of \studentdist\ distribution with $\dofs$ \\
 & & relative to Gaussian with same location/scale\\
\hline
$\decel$ &
  $\decel$ &
      deceleration parameter \\
$\decelhat$ &
  - &
      measured deceleration parameter \\
$\sigma_{\decel}$ &
  - &
      uncertainty on $\decel$ measurement \\
$\ddecel$ & &
      difference between local and cosmological $\decel$ \\
$\hubble$ &
  $\hubble$ &
      Hubble constant \\
$\hubblehat$ &
  - &
      measured Hubble constant \\
$\Sigma_{\rm cos}$ &
  - &
      covariance matrix of joint $\hubble-\decel$ constraints \\
$\dhubble$ & &
      difference between local and cosmological $\hubble$ \\
\hline
\end{tabular}
\end{table}


\section{Existing algorithms for estimating $\hubble$}
\label{section:existing_algorithms}

Two main attempts have been made at estimating $\hubble$ from the~\riess\ dataset: that used by \riess\ themselves (\sect{riess}); and the modification proposed by \cite{Cardona_etal:2016} (\sect{cardona}).  The method adopted by \citealt{Efstathiou:2014} and applied to the \cite{Riess_etal:2011} data differed from that of \riess\ primarily in regards outlier removal, but it has the same in statistical structure, so is effectively described in \sect{riess} as well; moreover, the updated method of \riess\ does not rely so heavily on the removal of outliers.  \cite{Zhang_etal:2017} has also recently addressed some issues of the outlier removal approach of \cite{Riess_etal:2011}, arguing that the freedom in the precise cuts made results in an increased systematic uncertainty; however, this algorithm is yet to be applied to the \riess\ dataset.


\subsection{Riess et al.\ (2016) approach}
\label{section:riess}

The original work of~\cite{Riess_etal:2009,Riess_etal:2011,Riess_etal:2016} casts the distance ladder as a linear regression problem, employing least-squares solutions to estimate the maximum-likelihood values of the parameters of interest and their covariance. To speed up the analysis, the Hubble flow SNe are processed independently from the Cepheids. The SN data are summarized by the intercept of their magnitude-log-redshift relation,
\begin{align}
a_x = \log_{10} \left\{ \frac{c\,z}{\rm km\,s^{-1}} \right. & \bigg[ 1 + \frac{1}{2}(1 - \decel) z - \nonumber\\
& \left.\left. \frac{1}{6}\left(1 - \decel - 3\decel ^ 2 + j_0\right) z^2 \right] \right\} - \frac{\msup}{5},
\end{align}
using a (potentially biased: see~\citealt{Kelly:2007,March_etal:2011,Mandel_etal:2016}) $\chi^2$-minimization formalism which also estimates the parameters of the Tripp relation, including the intrinsic scatter~\citep{Marriner_etal:2011}. The anchor and Cepheid-host data are input into a generalized least-squares (GLS) solution to Equations~\ref{equation:cepheid_pzl} and~\ref{equation:tripp} (and a handful of constraint equations) to estimate the standardized SN absolute magnitude. The Hubble constant is then given in terms of these two quantities as
\begin{equation}
\hubble = 10^{(\Mstds + 5a_x + 25) / 5}.
\end{equation}

There are a number of approximations and assumptions in this analysis that might affect the accurate estimation of the tails of the $\hubble$ posterior. Implicit in the GLS framework is the assumption that the data and quantities of interest are distributed normally and related linearly. As a result, measurements of anchor distances and parallaxes (which, in the case of the MASER measurements, are already non-Gaussian) must be converted to Gaussian uncertainties on distance moduli. Furthermore, though the framework produces Gaussian constraints on $\log_{10} H_0$, uncertainties on $H_0$ are quoted with symmetric error bars, implying a Gaussian posterior on $H_0$. With good data, the assumptions implicit in this framework should not bias the central value, but estimating tension between datasets requires precise knowledge of the tails of the posteriors, where deviations from Gaussianity are important. In this case, employing the `correct' log-normal posterior for $H_0$ instead of the Gaussian approximation {\em exacerbates} the tension between local measurements and the value extrapolated from {\it Planck}: the \citet{Planck_Int_XLVI:2016} value is 1.7 times less likely in a log-normal posterior than the equivalent Gaussian.

\begin{figure*}
\includegraphics[width=2\columnwidth]{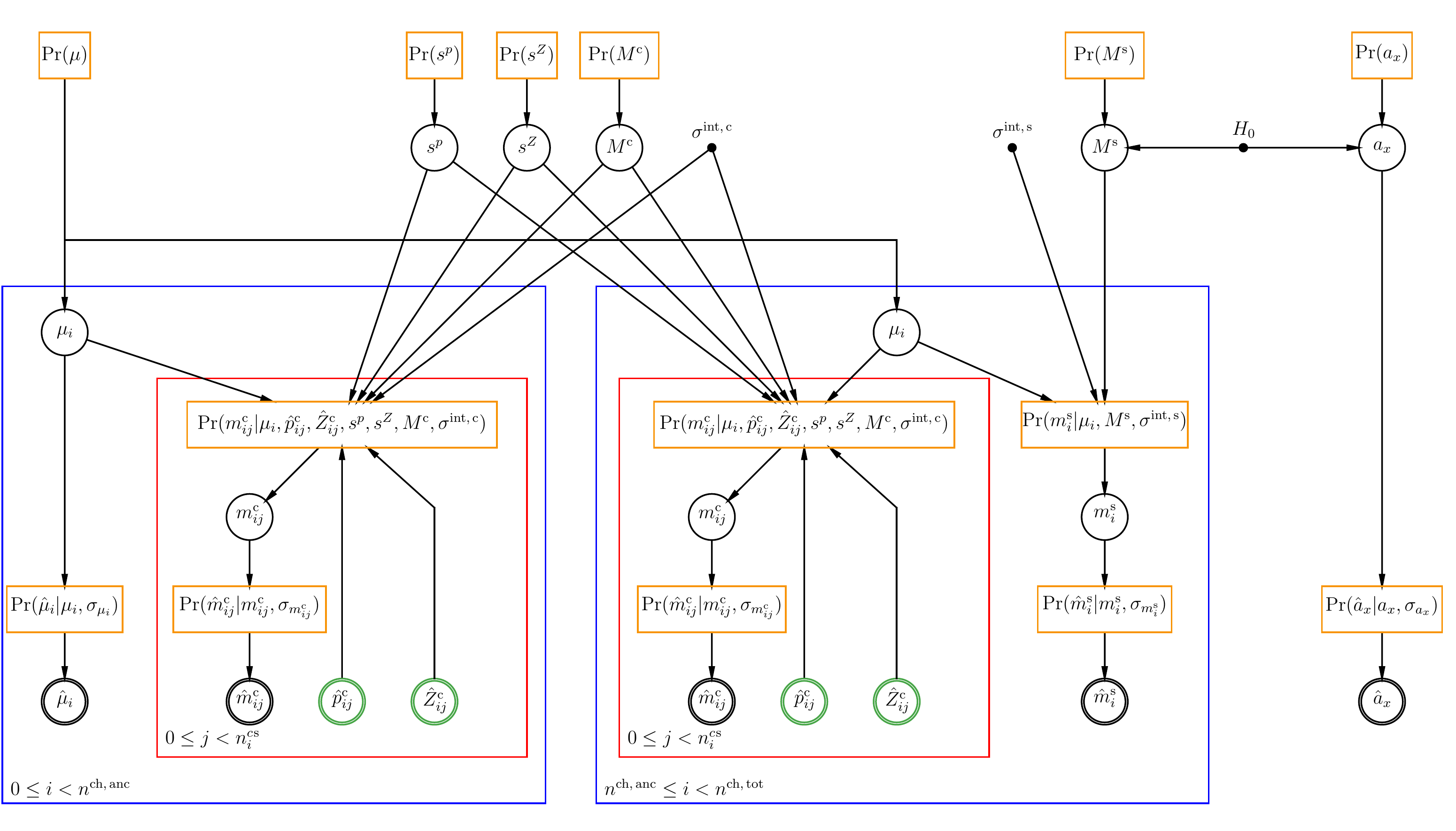}
\caption{The Bayesian network equivalent of the~\riess\ generalized least squares estimator for the analysis of the anchor, Cepheid and {\em compressed} SN data. Green double circles represent data measured without uncertainty, filled black circles indicate non-stochastic variables.}
\label{figure:network_r16}
\end{figure*}

The GLS framework is further restricted in its ability to cope with uncertainties in the independent variables and parameters affecting the observation covariance matrix. In the former case, one must either neglect uncertainties in certain variables or convert them into uncertainties on dependent variables assuming some set of fiducial parameter values. In the latter case, estimating parameters such as the Cepheids' intrinsic scatter (if desired) requires the analysis to be iterated to convergence. Iteration is also employed to remove Cepheid outliers via sigma clipping; again, this modifies the tails of joint posterior, whose accurate determination is critical to the estimation of tension between datasets. \cite{Efstathiou:2014} investigated modifications to this outlier rejection algorithm, replacing galaxy-by-galaxy rejection with a global version, though neither the core least-squares estimator nor the findings were significantly different.~\riess\ use supplementary Cepheid observations and simulations to improve targeted (as opposed to statistical) outlier identification, reducing the fraction of outliers considerably, but sigma clipping is still used to remove the 2--5\% of the Cepheid population deemed to be atypical; sigma clipping is also used to remove outliers in the SN dataset.

Finally, the compression of the Hubble flow SNe data into a single estimated intercept, $a_x$, introduces two further approximations. First, the covariance between the intercept and the low-redshift SNe's Tripp-relation parameters is neglected: the complete set of SNe should strictly be standardized simultaneously with the rest of the model in order to capture the correct correlation structure. Second, the deceleration parameter is fixed when calculating the intercept, and the degeneracy between $\decel$ and $\hubble$ is therefore neglected in the inference of $\hubble$. While \riess\ provide a heuristic estimate of the impact of varying $\decel$ on their uncertainties, there is also an associated shift in the best-fit value of $\hubble$ that is not accounted for. In addition, the value chosen for $\decel$ is derived from a $\Lambda$CDM-specific SN analysis~\citep{Betoule_etal:2014} that includes some SNe in common with those used by \riess; this introduces complicated dependence on the cosmological model that must be accounted for very carefully.  Combined with the fact that simultaneous inference of the Cepheid and SN parameters holds the promise for decreased uncertainties (due to shrinkage; see also~\citet{Zhang_etal:2017}), there is a strong motivation for developing a method that goes beyond the propagation of uncertainties in the form of parameter covariances.

The restrictions and approximations in the GLS solution can be highlighted by comparing its graphical representation (\fig{network_r16}) with that of the `complete' model (\fig{network_full}). Data that are assumed to be measured without uncertainty (in this case, the Cepheid periods and metallicities) are colour-coded green, and non-stochastic parameters -- that are either fixed ($\sigintc$ and $\sigints$) or determined by other parameters ($\hubble$)---are plotted with small filled circles. Note that the Cepheid-host SNe are characterized by only their magnitudes, whose uncertainties contain the errors on their stretches and colours. All conditional distributions in~\fig{network_r16} are taken to be Gaussians, and all population-level parameter priors, though not explicitly specified, are improper uniform distributions. As we shall see, all of these restrictions, approximations and assumptions can be removed or handled in a more generic fashion by employing a Bayesian hierarchical model.


\subsection{Cardona et al.\ (2016) approach}
\label{section:cardona}

\cite{Cardona_etal:2016} adopt a Bayesian approach to inferring the distance ladder parameters, exploring the full posterior distribution (rather than estimating uncertainties from the curvature at the posterior maximum, as is effectively done by \riess). They also introduce `hyper-parameters' to re-weight the contributions of each object to the likelihood, and allow the Cepheid intrinsic scatter to be sampled and marginalized (though the Hubble flow SNe data are still reduced to an intercept). This is, in essence, a Bayesian hierarchical model with pre-marginalization, although it is not explicitly described as such. The implicit network has the same structure as that used in~\riess\ (\fig{network_r16}), differing only through the sampling of $\sigintc$ from a log-uniform prior and the use of a Gaussian prior on $\slopeZ$.

\citet{Cardona_etal:2016} focus on improving the treatment of outliers in the Cepheid population. As an alternative to outlier rejection, the authors propose rescaling the magnitude uncertainty of the $j^{\rm th}$ Cepheid (in the $i^{\rm th}$ host) by a factor $\alpha_{ij} \sim \uniform(0, 1)$, such that $\sigma_{\mc}^2 \rightarrow \sigma_{\mc}^2 / \alpha_{ij}$; similar factors are included in the Cepheid-host SN magnitude likelihoods and anchor distance-modulus likelihoods. This rescaling factor is pre-marginalized, resulting in non-Gaussian likelihoods that have the same form as a student-t distribution with two degrees of freedom (\cf\ Eq.~\ref{equation:student}). Although boosting the tails of the likelihood to account for outliers is well-motivated, there is no particular reason that measurements should be rescaled individually, or that the prior on the rescaling should take its given form. It is perhaps better motivated to consider approaches that model outliers (or systematic errors) at a population level, for example by fitting multiple Cepheid populations in a mixture model (following, \eg,~\citealt{Hogg_etal:2010}), or using non-Gaussian likelihoods whose parameters are linked at the population level: either by error-rescaling factors common to measurements made on the same instrument or in similar epochs (\eg, \citealt{Lahav_etal:2000}) or, more simply, modifying the form of the scatter around the Cepheid P-L-Z relation. Finally, we note that the Cepheid catalogue released by~\riess\ and processed by~\citet{Cardona_etal:2016} has had outliers removed by sigma clipping, so the full power of this method to mitigate outliers in the \riess\ Cepheids has not yet been demonstrated.


\section{Bayesian parameter inference}
\label{section:model}

In statistical terms, the task of
inferring the Hubble constant
using the diverse astronomical data described in \sect{ladder}
is a conceptually straightforward case of parameter
estimation, albeit with 
a large number of nuisance parameters.
Bayesian inference provides a coherent formalism for 
this which can
incorporate all the relevant information that is available
while forcing all assumptions to be made explicit.

\subsection{The posterior distribution of \mbox{\boldmath$H_0$}}

The primary output of this calculation will be
the (posterior) probability distribution 
$\prob(\hubble, \nuisance | \data)$,
which encodes 
the full state of knowledge about
$\hubble$ and 
the other astrophysical parameters, $\nuisance$ 
(\ie, all the unobserved quantities 
that describe the anchors, the Cepheids, the SNe and their
host galaxies, and the cosmological model, 
as defined in \tabl{params})
given 
the available data, $\data$
(\ie, all the measured values for the various anchors, Cepheids and SNe,
again defined in \tabl{params}).
Bayes's theorem,
combined with the law of total probability,
then allows the posterior distribution to be written as 
\begin{equation}
\label{equation:bt}
\prob(\hubble, \nuisance | \data)
  = \frac{\prob(\hubble, \nuisance) \, 
  \prob(\data | \hubble, \nuisance)}
  {\int \prob(\hubble^\prime, \nuisance^\prime) \, 
  \prob(\data | \hubble^\prime, \nuisance^\prime) 
  \, \diff \hubble^\prime \, \diff \nuisance^\prime},
\end{equation}
where 
$\prob(\hubble, \nuisance)$ encodes any prior/external
information about $\hubble$ and $\nuisance$
and 
$\prob(\data | \hubble, \nuisance)$
is the probability of obtaining the measured
data given a set of parameter values (\ie, the likelihood).
The denominator in \eq{bt} is, in this parameter estimation context, 
just a normalizing constant which has no dependence on either
$\hubble$ or $\nuisance$; it is hence only the parameter-dependence
of the two terms in the numerator that is important.

The joint posterior distribution 
$\prob(\hubble, \nuisance | \data)$ is 
difficult to manipulate in its full form
as there are so many parameters,
but integrating 
over $\nuisance$ 
produces the
marginal posterior distribution of the Hubble constant,
\begin{equation}
\prob(\hubble | \data)
  = \frac
  {\int \prob(\hubble, \nuisance^{\prime\prime}) \, 
  \prob(\data | \hubble^{\prime\prime}, \nuisance^{\prime\prime}) 
  \, \diff \nuisance^{\prime\prime}}
  {\int \prob(\hubble^\prime, \nuisance^\prime) \, 
  \prob(\data | \hubble^\prime, \nuisance^\prime) 
  \, \diff \hubble^\prime \, \diff \nuisance^\prime}.
\end{equation}
Fortunately, the integrals in the numerator and denominator 
can be evaluated accurately provided only that it is 
possible to generate a large number of samples from 
the joint distribution $\prob(\hubble, \nuisance | \data)$.

\subsection{Bayesian hierarchical model}

The starting point for adapting this general formalism to the specific problem of estimating $\hubble$ is to list all of the quantities that are relevant to this problem, as already set out in \tabl{params}.
Global parameters (\eg\ those describing the Cepheid P-L-Z relationship) require prior distributions, which in all cases are taken to be as uninformative as possible: this is a data-rich problem in which it is the implications of the specific anchor, Cepheid and SN measurements that are of interest. The prior distributions adopted here are listed in~\tabl{relationships}.


\begin{table*}
\centering
\scriptsize
\caption{Probability distributions. The shorthand functions used derive from Equations~\ref{equation:cepheid_pzl}-\ref{equation:z2mu} and are defined at the foot of the table.}
\label{table:relationships}
\begin{tabular}{llrll}
\hline
distribution & form & process \\
\hline
$\prob(\mu)$ &
  $\uniform(5, 40)$ &
    prior on Cepheid/SN host distance moduli \\
$\prob(\slopep)$ &
  $\normal(\slopep; -5, 5^2)$ &
    prior on log-period slope of Cepheid luminosities \\
$\prob(\slopeZ)$ &
  $\normal(\slopeZ; 0, 5^2)$ &
    prior on log-metallicity slope of Cepheid luminosities \\
$\prob(\Mstdc)$ &
  $\normal(\Mstdc; 0, 20^2)$ &
    prior on standard Cepheid absolute magnitude \\
$\prob(\sigintc)$ &
  $\normal(\sigintc; 0.1, 0.2^2) \times \Theta(\sigintc-0.01) \times \Theta(3-\sigintc)$ &
    prior on intrinsic scatter in Cepheid Leavitt Law \\
$\prob(\dmcal)$ &
  $\normal(\dmcal; 0, 0.03^2)$ &
    prior on ground-to-space magnitude offset \\
$\prob(\Mstds)$ &
  $\normal(\Mstds;-20, 10^2)$ &
    prior on standard SN absolute magnitude \\
$\prob(\sigints)$ &
  $\normal(\sigints;0.1, 0.2^2) \times \Theta(\sigints-0.01) \times \Theta(3-\sigints)$ &
    prior on intrinsic scatter of SNe around Tripp relation \\
$\prob(\xsalt)$ &
  $\normal(\xsalt;0, 2^2)$ &
    prior on SN stretches \\
$\prob(\csalt)$ &
  $\normal(\csalt;0, 2^2)$ &
    prior on SN colours \\
$\prob(\alphas)$ &
  $\normal(\alphas;-0.1, 0.5^2)$ &
    prior on SN stretch coefficient \\
$\prob(\betas)$ &
  $\normal(\betas;3, 3^2)$ &
    prior on SN colour coefficient \\
$\prob(z)$ &
  $\uniform(0.01, 0.15)$ &
    prior on SN redshift distribution \\
$\prob(\hubble)$ &
  $\normal(\hubble;70, 20^2)$ &
    prior on Hubble Constant\\
$\prob(\decel)$ &
  $\normal(\decel;-0.5, 1^2) \times \Theta(\decel+5) \times \Theta(1-\decel)$ &
    prior on deceleration parameter \\
$\prob(\mc|\mu_i,\phatc,\Zhatc,\slopep,\slopeZ,\Mstdc,\sigintc)$ &
  $\normal(\mc; \mu_i + M^{\rm c}_{ij}(\Mstdc,\phatc,\Zhatc,\slopep,\slopeZ) [+ \dmcal], (\sigintc)^2) $&
    intrinsic scatter about Cepheid Leavitt Law \\
$\prob(\msup|\mu_i,\Mstds,\sigints)$ &
  $\normal(\msup;\mu_i + \Mstds,(\sigints)^2)$ &
    intrinsic scatter about Cepheid-host distance modulus \\
$\prob(\msup|\zs,\xsalt_i,\csalt_i,\Mstds,\alphas,\betas,\hubble,\decel,\sigints)$ &
  $\normal(\msup; \mu_i(\zs,\hubble,\decel) + M_i^{\rm s}(\Mstds,\xsalt_i,\csalt_i,\alphas,\betas), (\sigints)^2)$ &
    intrinsic scatter about Hubble flow SN Tripp Relation \\
$\prob(\hat{d}_i|\mu_i,\sigma_{d_i})$ &
  $\normal(\hat{d}_i;10^{(\mu_i + 5) / 5},\sigma_{d_i}^2)$ &
    anchor distance measurement \\
$\prob(\hat{\pi}_i|\mu_i,\sigma_{\pi_i})$ &
  $\normal(\hat{\pi}_i;10^{-(\mu_i + \dMLK + 5) / 5},\sigma_{\pi_i}^2)$ &
    anchor parallax measurement \\
$\prob(\mhatc|\mc,\sigma_{\mc})$ &
  $\normal(\mhatc;\mc,\sigma_{\mc}^2)$ &
    measurement of Cepheid apparent magnitude \\
$\prob(\mhats|\msup,\sigma_{\msup})$ &
  $\normal(\mhats;\msup,\sigma_{\msup}^2)$ &
    measurement of Cepheid-host SN apparent magnitude \\
$\prob(\mhats,\xsalthats,\csalthats|\msup,\xsalts,\csalts,\Sigma_i)$ &
  $\normal(\{\mhats,\xsalthats,\csalthats\};\{\msup,\xsalts,\csalts\},\Sigma_i)$ &
    correlated SALT-2 estimates of SNe observables \\
$\prob(\zhats|\zs,\sigma_{\zs})$ &
  $\normal(\zhats; \zs, \sigma_{\zs}^2)$ &
    SN redshift measurement \\
$\prob(\hat{q}_0|\decel,\sigma_{\decel})$ &
  $\normal(\hat{q}_0; -0.5575, 0.051^2)$ &
    \citet{Betoule_etal:2014} estimate of deceleration parameter \\
\hline
$\prob(\tpeakc)$ &
  $\uniform(0, 1)$ &
    prior on degrees of freedom in Cepheid scatter \\
$\prob(\tpeaks)$ &
  $\uniform(0, 1)$ &
    prior on degrees of freedom in SN scatter \\
$\prob(\mc|\mu_i,\phatc,\Zhatc,\slopep,\slopeZ,\Mstdc,\sigintc)$ &
  ${\rm T}(\mc; \dofc, \mu_i + M^{\rm c}_{ij}(\phatc,\Zhatc,\slopep,\slopeZ) [+ \dmcal], (\sigintc)^2) $&
    heavy-tailed Cepheid Leavitt Law \\
$\prob(\msup|\mu_i,\Mstds,\sigints)$ &
  ${\rm T}(\msup;\dofs,\mu_i + \Mstds,(\sigints)^2)$ &
    Cepheid-host distance modulus \\
$\prob(\msup|\zhats,\xsalt_i,\csalt_i,\Mstds,\alphas,\betas,\hubble,\decel,\sigints)$ &
  ${\rm T}(\msup; \dofs, \mu_i(\zhats,\hubble,\decel) + M_i^{\rm s}(\xsalt_i,\csalt_i,\alphas,\betas), (\sigints)^2)$ &
    heavy-tailed Hubble flow SN Tripp relation \\
\hline
$\prob(\dhubble)$ &
  $\normal(\dhubble;0, 6^2)$ &
    prior on difference between local and cosmological $\hubble$\\
$\prob(\ddecel)$ &
  $\normal(\ddecel;0, 0.5^2)$ &
    prior on difference between local and cosmological $\decel$ \\
\hline
\multicolumn{3}{p{\columnwidth}}{Notes:} \\
\multicolumn{3}{p{\columnwidth}}{$M^{\rm c}_{ij}(\Mstdc,\phatc,\Zhatc,\slopep,\slopeZ) \equiv \Mstdc + \slopep \log_{10} \phatc  + \slopeZ \log_{10} \Zhatc$} \\
\multicolumn{3}{p{\columnwidth}}{$M_i^{\rm s}(\Mstds,\xsalt_i,\csalt_i,\alphas,\betas) \equiv \Mstds + \alphas \xsalt_i + \betas \csalt_i$} \\
\multicolumn{3}{p{\columnwidth}}{$\mu_i(\zs,\hubble,\decel) \equiv 5 \log_{10} \left\{ \frac{c\,\zs}{\hubble {\rm pc}} \left[ 1 + \frac{1}{2}(1 - \decel) \zs - \frac{1}{6}\left(2 - \decel - 3\decel ^ 2\right) \left(\zs \right)^2 \right] \right\} + 10$}
\end{tabular}
\end{table*}

The next step is to make explicit all the direct relationships between the quantities defined in \tabl{params}, as it is these which combine to determine the form of the likelihood, $\prob(\data | \hubble, \nuisance)$. 
The full model is illustrated in \fig{network}, 
from which it is clear that most pairs of quantities are 
not linked directly, with a multi-layered structure that is characteristic of a Bayesian hierarchical model (BHM).
This structure notwithstanding, all the data and parameters 
are linked indirectly, which is why a joint analysis 
of the anchor, Cepheid and SN data is required to 
obtain a posterior distribution for $\hubble$ that incorporates all relevant sources of uncertainty.
Perhaps most interesting is the longest chain of links,
from the nearest anchor distance, $\hat{d}_0$, 
through the entire network to $\hubble$.
Any uncertainty in $\hat{d}_0$
will hence increase the uncertainty on $\hubble$, although 
it is only this qualitative statement that can be made at this point:
the magnitude and form of this dependence depends on the nature of the links.

Each of the inter-parameter links in \fig{network} is encoded by 
a probability distribution relating the value of the 
quantity at the head of the arrows to the quantities at the tails.
(In some cases a deterministic approximation can be made;
this can still be cast in a probablistic form by using 
delta-function distributions.)
It is at this stage that many of the implicit model assumptions -- most 
obviously that certain distributions are normal,
rather than having heavier tails -- must be made explicit.
The full list of relationships used here is given in \tabl{relationships} along
with the global-parameter priors; they are also indicated within orange
rectangles in \fig{network}.

The BHM has a number of important features that ensure a number of potential statistical problems are dealt with automatically.
We consider the complete dataset simultaneously, fitting the SN
Tripp relation concurrently with the Cepheid Leavitt Law\footnote{The blind analysis pipeline of~\cite{Zhang_etal:2017}, published as this manuscript was finalized, also performs a global fit for the combined Cepheid and SNe data.}, taking account of redshift 
measurement errors and peculiar velocity uncertainties, correlations between 
the estimated magnitudes, stretches and colours of the Hubble flow SNe, and uncertainty in the 
measured value of the deceleration parameter, $\decel$. Fitting the SN population using a BHM removes the bias injected by the $\chi^2$-minimization routines popular in the literature (see~\citealt{Kelly:2007,March_etal:2011,Mandel_etal:2016} for discussion). In the vanilla 
version of the BHM, the scatter about the Cepheid Leavitt and SN Tripp relations is taken to be Gaussian. This is a significant assumption, given 
that considerable pre-processing is typically performed to remove 
outliers from the Cepheid and SN populations. 
Though~\riess\ use supplementary Cepheid observations and simulations 
to reduce the number of Cepheid outliers, sigma clipping is still employed 
in both the Cepheid and SN selection. As such, we also consider
a modified BHM in which the Cepheid and SN scatters are modeled 
using a student-t distribution with degrees of freedom inferred from the data.\footnote{An  
alternative approach would be to model the contaminated Cepheids and SNe as multiple-component 
mixtures, each with their own Gaussian scatter and absolute magnitude. We have 
implemented such a model, but our selected sampler can not reliably sample the resulting 
multimodal posterior and its strict relabeling degeneracies.}
Varying the degrees of freedom interpolates between Gaussian (in the limit 
of infinite degrees of freedom) and significantly heavy-tailed 
scatter distributions, allowing the data to indicate whether
outliers are present in the Cepheids and/or SNe. We use a bespoke prior 
for these parameters to allow the sampler to explore distributions 
with a broad range of kurtosis values. The kurtosis of the \studentdist\ distribution is not defined over the full range of allowed degrees of freedom, so we instead use the ratio of the peak density of the \studentdist\ to that of a Gaussian with the same location and scale, denoted $\tpeak$. We adopt a prior that is uniform in $\tpeak$, the derivation of which is covered in detail in Appendix~\ref{section:dof_prior}. Provided the outliers are well characterized by the model, this approach has two distinct advantages over sigma clipping: first, hard-earned but discrepant data are not discarded but instead appropriately down-weighted before use; second, no likelihoods are artificially truncated, allowing accurate and robust evaluation of the tails of the $\hubble$ posterior.

In the basic version of the BHM, all data, including geometric anchor measurements, are taken to be observed with Gaussian uncertainties. This departs from previous work, which assumes the distance {\em moduli} (which are not directly measured) are subject to Gaussian uncertainties. As the shape of the anchor likelihood(s) propagates through to the inference on $\hubble$, we would expect our model to yield a more appropriate estimate of the tail of the $\hubble$ posterior. Using distances and parallaxes as data rather than distance moduli means we can also sample from non-Gaussian anchor likelihoods if merited: for example, the MASER distance constraint derived from a model of the MASER motion has a mildly non-Gaussian form~\citep{Humphreys_etal:2013,Riess_etal:2016}. Approximating the MASER distance posterior as a three-component Gaussian mixture, we can estimate the impact of this non-Gaussian anchor measurement on $\hubble$. More exciting is to exploit this structure to connect to other BHMs performing, for example, more accurate inference of parallax measurements~\citep{Sesar_etal:2016}, though this is left to 
future work. 

The BHM described above is realistic, but there are some aspects of it which could be improved or extended if the data demanded it.
First, as~\riess\ do not provide uncertainties on the Cepheid periods and metallicities, we are forced to model the distributions between the measured and true periods and metallicities as delta functions. Second, in order to minimize deviations from the~\riess\ dataset we retain their model of the Cepheid-host SNe magnitudes, i.e., they are assumed to have been pre-corrected based on their stretch and colour, with the uncertainties on stretch and colour propagated to the magnitude; in effect, we partially decouple the Cepheid-host and Hubble flow SNe.

\begin{landscape}
\begin{figure}
\includegraphics[width=\columnwidth]{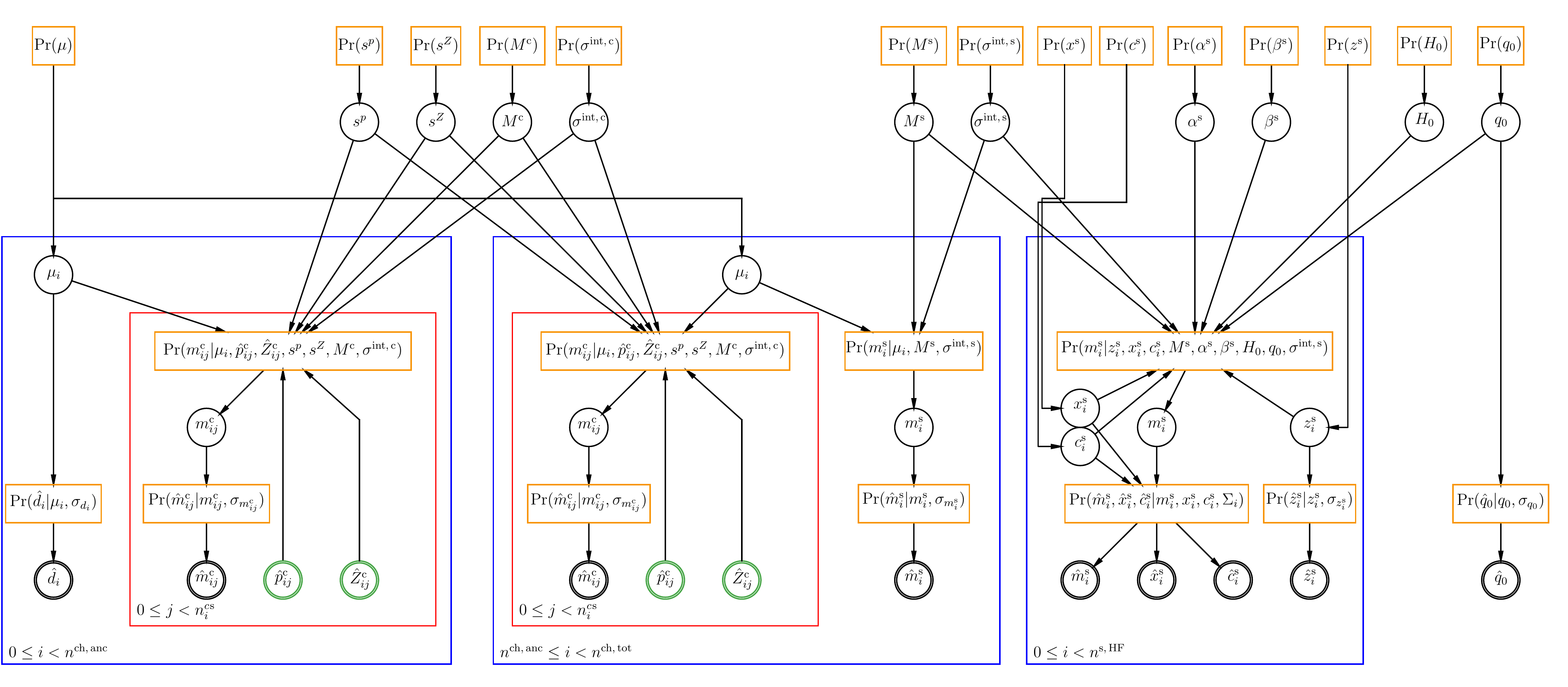}
\caption{The Bayesian network sampled from in this study. This is somewhat simplified from the full treatment to allow the~\riess\ data to be processed.}
\label{figure:network}
\end{figure}
\end{landscape}


\subsection{Hamiltonian Monte Carlo sampling scheme}
\label{section:hmc}

Having set out the form of the model and the distributions defining its links, we are faced with the choice of how to sample from it to estimate the joint posterior. As stated earlier, the sparse structure of the model suggests Gibbs sampling could be well suited to the task; however, given the large number of parameters involved, we instead employ Hamiltonian Monte Carlo sampling~\citep{Duane_etal:1987,Neal:2012}. Hamiltonian (or hybrid) Monte Carlo (HMC), is a sampling method designed to avoid the random walk behaviour which hinders other Markov chain Monte Carlo (MCMC) techniques in high-dimension or highly degenerate settings. It operates by simulating a physical system governed by Hamiltonian dynamics, modeling samples as particles moving through parameter space with potential energy equal to the negative logarithm of the posterior. This requires a momentum variable to be sampled for each particle: at each iteration, a new momentum is proposed and used to move the particle through parameter space using a time-discretized form of Hamilton's equations. The new position is then accepted in a Metropolis accept/reject step based on the ratio of the evolved and initial Hamiltonians. Though sampling the momenta doubles the number of unknown variables in the sampling problem, modifying the particles' kinetic energies allows the sampler to propose large steps in parameter space, enabling rapid  posterior exploration.

Sampling efficiently with HMC requires a significant amount of fine-tuning. Calculating the Hamiltonian requires evaluation of the derivative of the log posterior with respect to the parameters of interest, a potentially error-prone operation when no analytical form exists. The particles' dynamics are evolved by solving Hamilton's equations with time discretized: this requires the number of discrete time steps and their duration to be specified. Furthermore, in defining the momentum of each particle a mass must be chosen. Much of the art of HMC rests in tuning these parameters. We use the Stan package~\citep{pystan} to optimize these choices for us. Provided with a description of the model in its scripting language, Stan tunes the required parameters on the fly and uses auto-differentiation to calculate machine-precision posterior derivatives. Translating the model into Stan's scripting language~\citep{stan_manual} is (modulo modifications for optimization) as simple as specifying the analytic form of every link between a probability distribution and a variable, latent or otherwise, in Figure~\ref{figure:network}, as provided in Table~\ref{table:relationships}. Where reasonable, priors have been chosen to to be weakly informative normal distributions to aid sampling: Stan works best without sharp limits. The Stan model code and Python driver used in this analysis are publicly available at \url{https://github.com/sfeeney/hh0}.


\section{Data selection}
\label{section:data}

\riess\ produced $H_0$ estimates for a comprehensive set of anchor combinations, further varying the Cepheid and SN counts by changing the outlier-rejection method and allowed redshift range, respectively. Here, for clarity, we compare to two \riess\ settings. In the first, simple setting, we consider a single distance anchor: the MASER in NGC 4258. We also process \riess's preferred anchor combination, employing three anchors: the NGC 4258 MASER, the LMC detached-eclipsing-binary distance and the set of 15 Milky Way Cepheids with parallax measurements. In both cases, we consider 19 galaxies containing a population of Cepheids and a Type Ia SN, and Hubble flow SNe in the redshift range $0.0233 < z < 0.15$. We bolster the Cepheid sample by including the set of Cepheids measured in M31~\citep{Riess_etal:2012,Kodric_etal:2015,Wagner-Kaiser_etal:2015}, which has not hosted an observed Type Ia SN. This calibration-only Cepheid host is not depicted in~\fig{network}, but appears in the model as a simplified version of a Cepheid/SN host. In total, the~\riess\ data contain 1486 and 2276 Cepheids in the one- and three-anchor cases respectively; note that these have already had outliers removed via sigma clipping on a galaxy-by-galaxy basis, limiting any investigation into alternative outlier-mitigation mechanisms.

The Hubble flow SNe are sourced from catalogues compiled by~\citet{Riess_etal:1999,Jha_etal:2006,Hicken_etal:2009a,Hicken_etal:2012,Guy_etal:2010,Stritzinger_etal:2011,Rest_etal:2014} and~\cite{Sako_etal:2014}, and 
calibrated according to~\cite{Scolnic_etal:2015} and~\cite{Scolnic_Kessler:2016}\footnote{Available for download from \url{http://kicp.uchicago.edu/~dscolnic/Supercal/supercal_vH0.fitres}.}. Cuts are applied matching those used by~\riess, namely: 
\begin{itemize}
\item
SN colors are required to fall in the range $|\csalt| < 0.3$; 
\item
Stretches must be $|\xsalt| < 3$ and estimated with an uncertainty of $< 1.5$;
\item
The light curve must have a fit probability of greater than 0.001, 
a peak time with uncertainty less than two days, 
and a peak magnitude with uncertainty less than 0.2 magnitudes.
\end{itemize}
We apply an additional cut, requiring that the covariance matrix, $\Sigma_i$, of the light curve parameters be positive definite. After cutting, \riess\ reject SN outliers via a $\chi^2$-minimization algorithm sketched out in~\cite{Marriner_etal:2011}. We analyze both the clean and contaminated catalogues in this work, considering 214 and 229 SNe, respectively.


\section{Analysis of simulated data}
\label{section:sim}

In order to isolate the statistical aspects of this issue it is instructive to work with simulated data, in which the complicated real world issues of systematics and outliers can be ignored. We construct our simulations to match the data selections discussed above, modeling both the one- and three-anchor settings using the same distances and uncertainties; for one-anchor simulations the anchor is therefore the MASER distance to NGC 4258; the three-anchor case contains a mixture of parallax and distance measurements. In addition to the anchors, we simulate 20 Cepheid hosts, 19 containing a SN, matching Cepheid counts to~\riess\ and taking the true host distance moduli to be the values estimated by~\riess's preferred analysis~\citep[see][Table 5]{Riess_etal:2016}.

The priors from which simulated Cepheid and SN observables are drawn are chosen for simplicity rather than to reproduce the~\riess\ dataset with perfect fidelity. The Cepheids' periods are taken as log-distributed in the range $5 < \phatc < 60$ days, and their metallicities as Gaussian-distributed with mean and standard deviation matching that of the~\riess\ sample (8.86 and 0.153, respectively)). Following~\riess, we assume the periods and metallicities have no uncertainties, though modeling such uncertainties is a simple extension to the hierarchical model. We adopt the following values for the parameters of the Cepheid P-L-Z relation ($\{\slopep,\slopeZ,\Mstdc\} = \{-3.05, -0.25, -3.09\}$), which is used to define the Cepheid absolute magnitudes. Intrinsic scatter in the relation is injected by drawing random deviates from the appropriate distribution: a zero-mean Gaussian with $\sigintc = 0.065$ initially; a student-t distribution with $\dofc=2$ when generating outliers. Finally, simulated Cepheid apparent magnitudes are produced by combining the absolute magnitudes with host distance moduli and Gaussian measurement uncertainties ($\sigma_{\mc} = 0.276$ to match that of the~\riess\ sample).

The nearby and Hubble flow SNe are simulated slightly differently, reflecting the different observables one must model to match the~\riess\ data. The Cepheid-host SNe are assigned apparent magnitudes using their host distance moduli assuming a SN absolute magnitude of $\Mstds = -19.2$, then noisily `observed' with Gaussian magnitude uncertainties with $\sigma_{\msup} = 0.064$. Zero-point offsets of 0.01 magnitudes are added to selected hosts' Cepheids and SNe to match~\riess\ observations. Redshifts for the Hubble flow SNe are drawn uniformly in the range selected by~\riess\ ($0.0233 < z < 0.15$), before adding Gaussian uncertainties due to measurement ($\sigma=10^{-5}$) and peculiar velocities ($\sigma=250\,{\rm km s}^{-1}/ c$). True colour and stretch parameters for each SN are drawn from Gaussian distributions with moments matching those of the sample used by~\riess; these are then combined with the true redshifts, SN absolute magnitude and random deviates drawn from an intrinsic scatter distribution (either a zero-mean normal distribution with $\sigints = 0.1$ or, if outliers are desired, a student-t distribution with $\dofs=2$) to generate true apparent magnitudes. This process requires the SN colour and stretch coefficients to be specified, along with the Hubble constant and the deceleration parameter. We take the SN coefficients to be 3.1 and -0.14 (see, \eg,~\citealt{Scolnic_Kessler:2016}), respectively, and the deceleration parameter to be -0.5575~\citep{Betoule_etal:2014}; the ground truth $\hubble$ value varies with the setting considered. The final simulated SN catalogue is produced by generating correlated Gaussian uncertainties on the SN apparent magnitudes, colours and stretches using the average correlations of the~\riess\ SN dataset, mimicking the effects of estimating these parameters from SN lightcurves with using SALT-2~\citep{Guy_etal:2007}.


\subsection{Simple model tests}
\label{section:simple_sim}

\begin{figure*}
\includegraphics[width=0.73\figwidth]{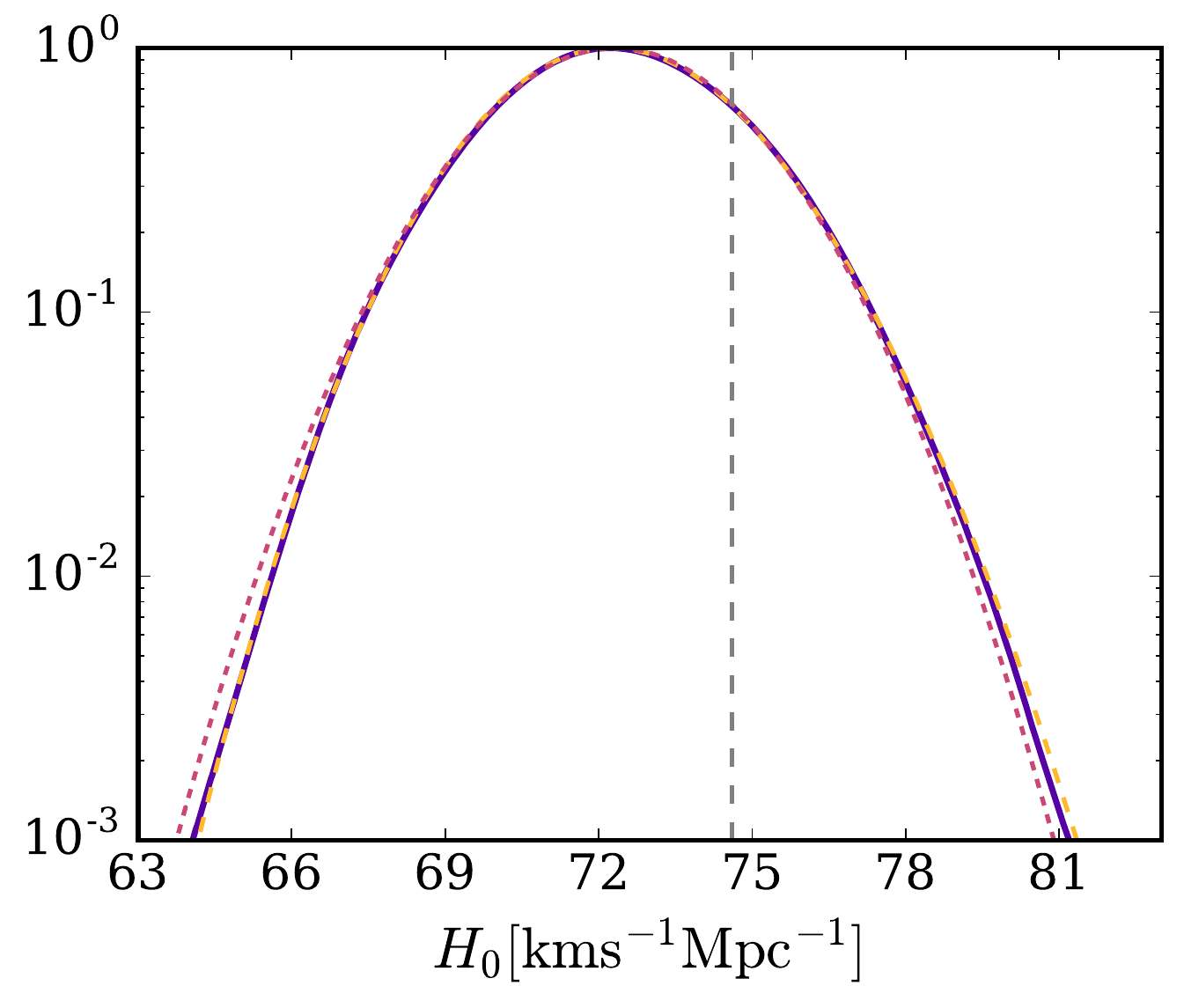}
\includegraphics[width=0.73\figwidth]{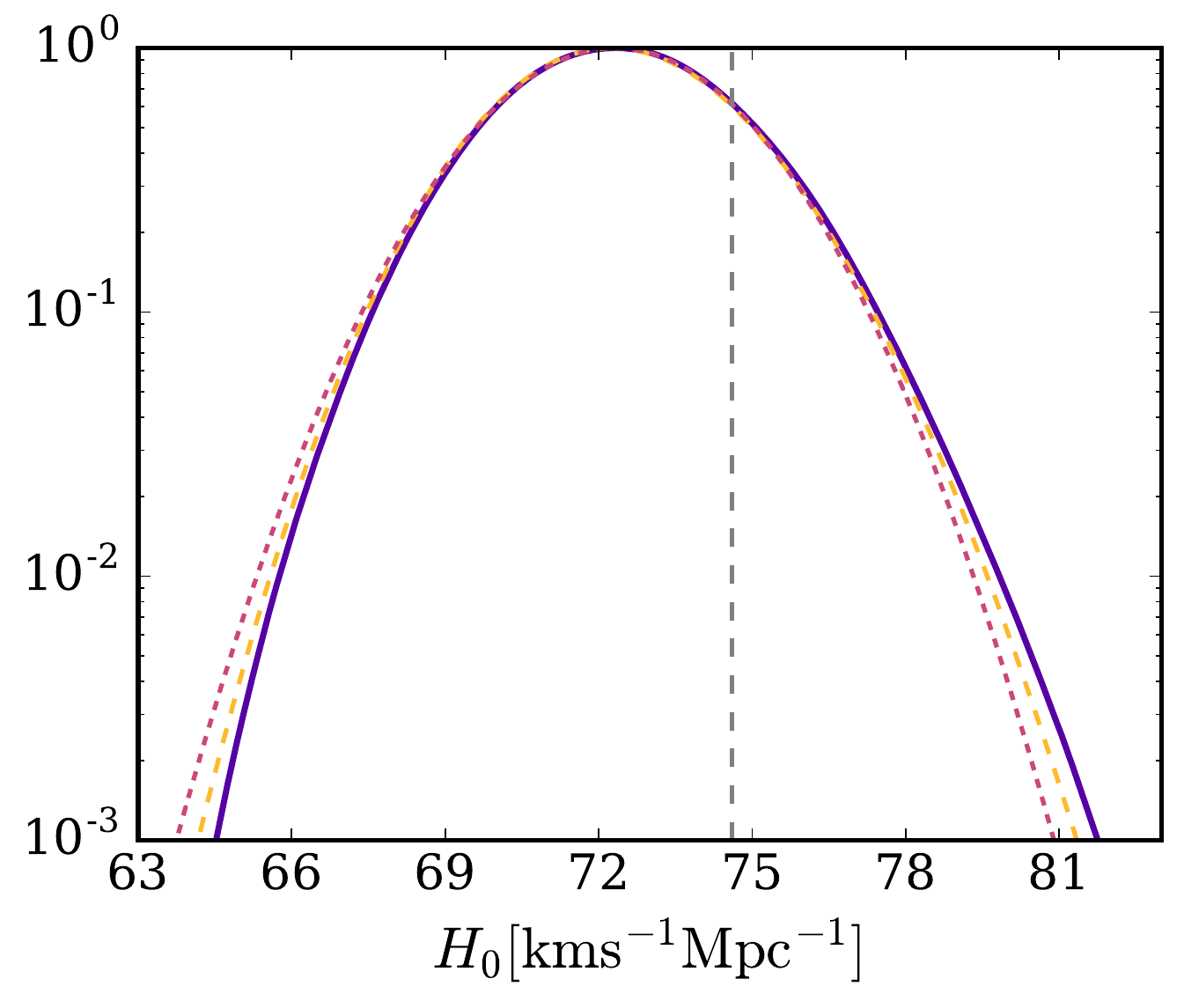}
\includegraphics[width=0.73\figwidth]{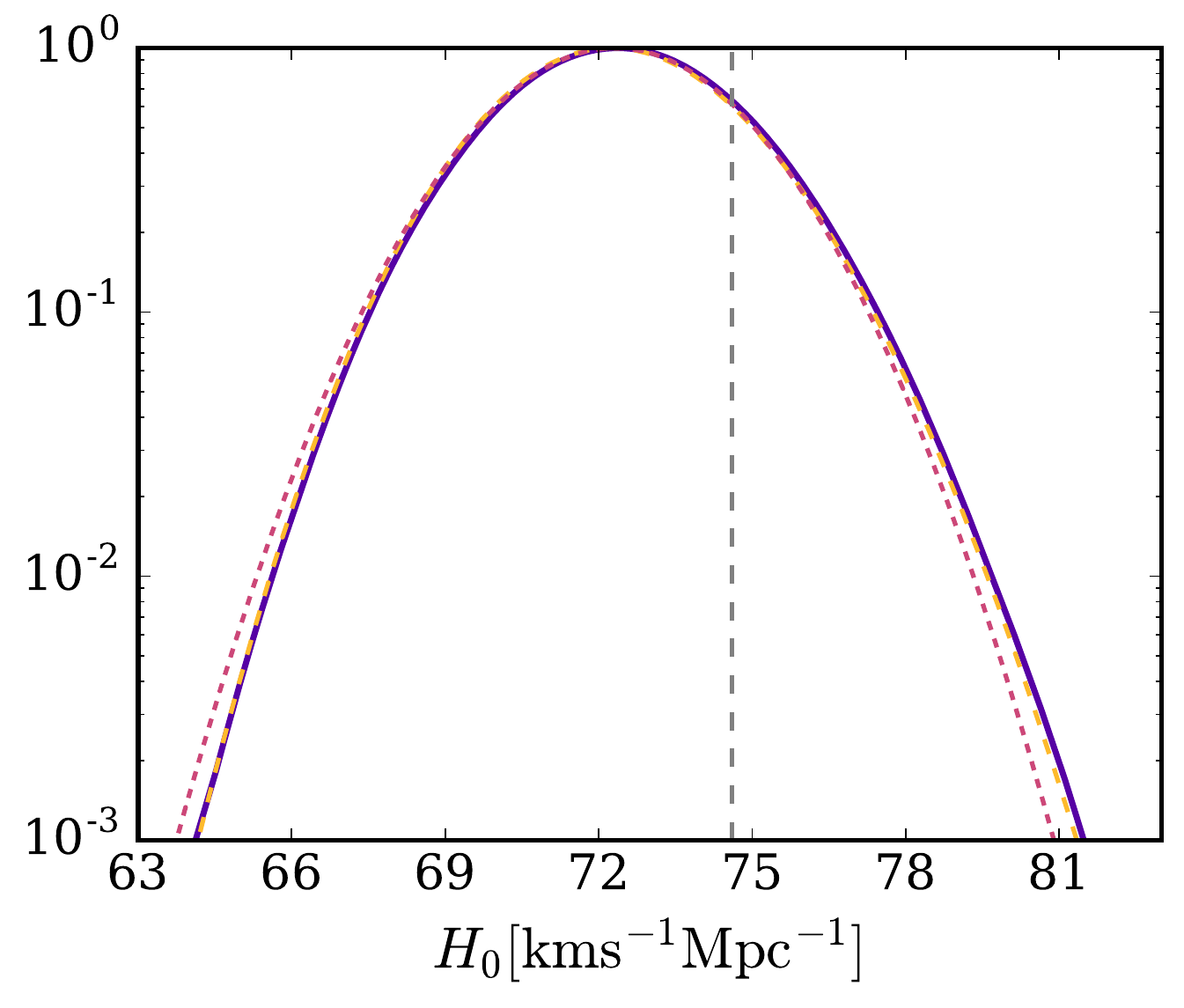}
\caption{Posteriors on $\hubble$ for one-anchor simulation. Purple posteriors are produced by BHMs sampling an anchor likelihood that is Gaussian in distance modulus (left) and distance (centre), and non-Gaussian in distance (right). Log-normal posteriors from the~\protect\riess\ GLS method are plotted as yellow long-dashed lines; approximated Gaussian posteriors (as typically quoted in the literature) are plotted as pink short-dashed lines; these curves are the same in each panel. Overlaid as a grey line is the ground truth $\hubble$ value for the simulation.}
\label{figure:simple_sim_h_0_comp}
\end{figure*}

We begin with an investigation into how the choice of anchor distance parametrization affects the final posterior on $\hubble$. As direct comparison to the GLS method is particularly instructive in this setting, we simplify the model following~\riess, replacing the simulated Hubble flow SN portion of the BHM with a single parameter, $a_x \sim \normal(0.71273, 0.00176^2)$. Initially, as a basic test of the sampler's performance we further simplify the problem, modifying the BHM to sample from Gaussian uncertainties on the anchor distance {\em moduli}, and converting the simulated anchor measurements accordingly. In this setting, we should expect the BHM and GLS posteriors to match identically, at least in the limit of infinite samples.

The $\hubble$ posterior estimated from a single-anchor simulation by the simplified BHM (using four chains each comprising 50,000 samples) is shown as a purple solid line in the left panel of \fig{simple_sim_h_0_comp}. Overlaid are the log-normal posterior as estimated by the GLS solution (dashed yellow) and the approximated Gaussian constraint as typically quoted (short-dashed pink). The agreement between the sampled posterior and the analytic log-normal distribution produced by the GLS is excellent, with only slight deviations in the high $\hubble$ tail; in all cases, the ground truth value of $\hubble$ is within the 68\% credible interval. The bias on the credible intervals injected by approximating the posterior as Gaussian is small but clear, with an overall shift of probability density to lower values of $\hubble$.

Reverting the BHM to use an anchor uncertainty that is Gaussian in {\em distance} and processing the same simulation skews the $\hubble$ posterior to even higher values, as shown in the centre panel of \fig{simple_sim_h_0_comp}. This is a simple consequence of the non-linearity of the $d \rightarrow \mu$ transform: the distance modulus likelihood becomes skewed towards lower values, which in turn boosts the high $\hubble$ tail. Replacing the anchor distance likelihood with our fit to the non-Gaussian MASER distance posterior of~\riess\ has the opposite effect, pushing the sampled posterior back into agreement with the log-normal GLS posterior (\fig{simple_sim_h_0_comp} right panel). This coincidence arises as the MASER distance posterior is close to, but not exactly, log-normal in distance. As asserted earlier, the posterior peak is effectively independent of the method and model used. From these tests, it is clear that in order to accurately resolve the tails of the $\hubble$ posterior it is important to model the anchor uncertainties (and, on a simpler level, the logarithmic nature of the GLS constraints) correctly.


\subsection{Outlier tests}
\label{section:outlier_sim}

\begin{figure}
\includegraphics[width=\figwidth]{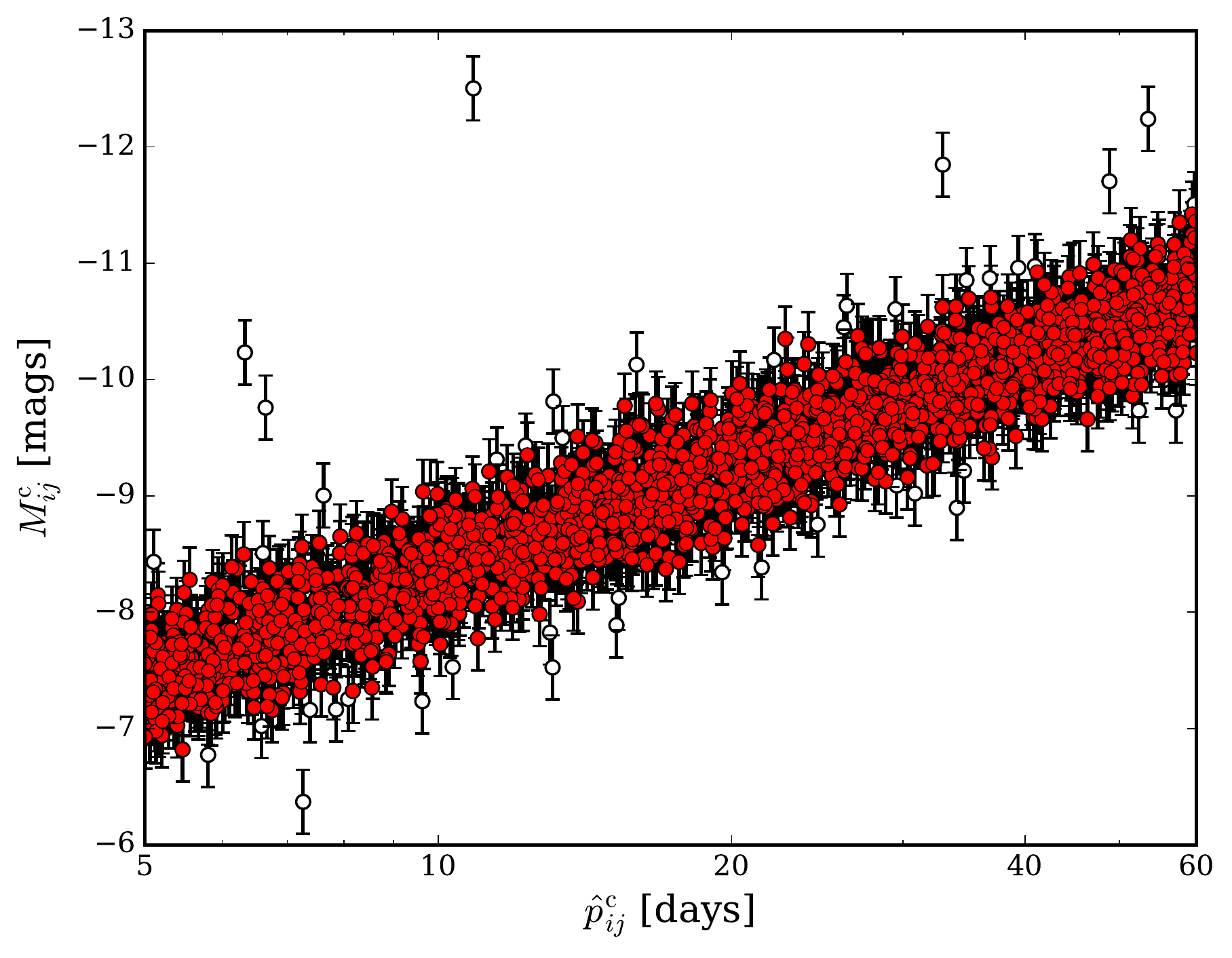}
\caption{Cepheid period-luminosity relation for simulation with heavy-tailed intrinsic scatter. Cepheids rejected by the~\protect\riess\ sigma-clipping algorithm are plotted with open circles.}
\label{figure:ht_ceph_sim}
\end{figure}

Having established good agreement between the simplified BHM and the GLS estimator, we now investigate the full complexity of the hierarchical model by fitting simulations containing Cepheid and SN outliers. Here, we consider the~\riess\ three-anchor setting, simulating and sampling the SNe and adding intrinsic scatter to the Cepheid and SNe magnitudes drawn from a \studentdist\ distribution (with $\dofc=\dofs=2$), mimicking the heavy tails injected by outlying observations. The resulting sample of Cepheids is plotted in~\fig{ht_ceph_sim}, with stars deemed as outliers by the~\riess\ sigma-clipping algorithm indicated. Even though Stan samples a total of 3,213 parameters in this setting, generating four chains of 50,000 samples (plus another 50,000 `warmup' samples) each takes roughly 8 hours on four 3.4 GHz Intel Xeon cores. From these chains, Stan estimates that roughly 30,000 independent samples have been taken from the $\hubble$ posterior, 20,000 from $\dofs$ and 2100 from $\dofc$. The SN scatter distribution is better sampled as the SN intrinsic scatter scale, $\sigints$, is comparable to the observation uncertainties; the Cepheid observation uncertainties tend to swamp the intrinsic scatter. The resulting marginalized posteriors for $\hubble$, along with the Student shape parameters $\tpeakc$ and $\tpeaks$, are plotted in the left panel of~\fig{ht_triangle_sim}. The $\tpeak=0$ limit corresponds to extremely heavy-tailed distributions ($\nu \rightarrow 0$), and $\tpeak=1$ corresponds to Gaussian intrinsic scatter ($\nu \rightarrow \infty$). The posteriors on $\tpeakc$ and $\tpeaks$ are clearly peaked away from one, and so the BHM correctly reflects the fact that the intrinsic scatter distributions are non-Gaussian in this case; we note that this is, of course, a function of the noise level and number of sources. For all parameters shown in~\fig{ht_triangle_sim} the ground truth values are well within the 68\% credible intervals; the largest discrepancy found for a population-level parameter is for $\slopep$, whose peak is $\sim1.5-\sigma$ from the ground truth.

\begin{figure*}
\includegraphics[width=\figwidth]{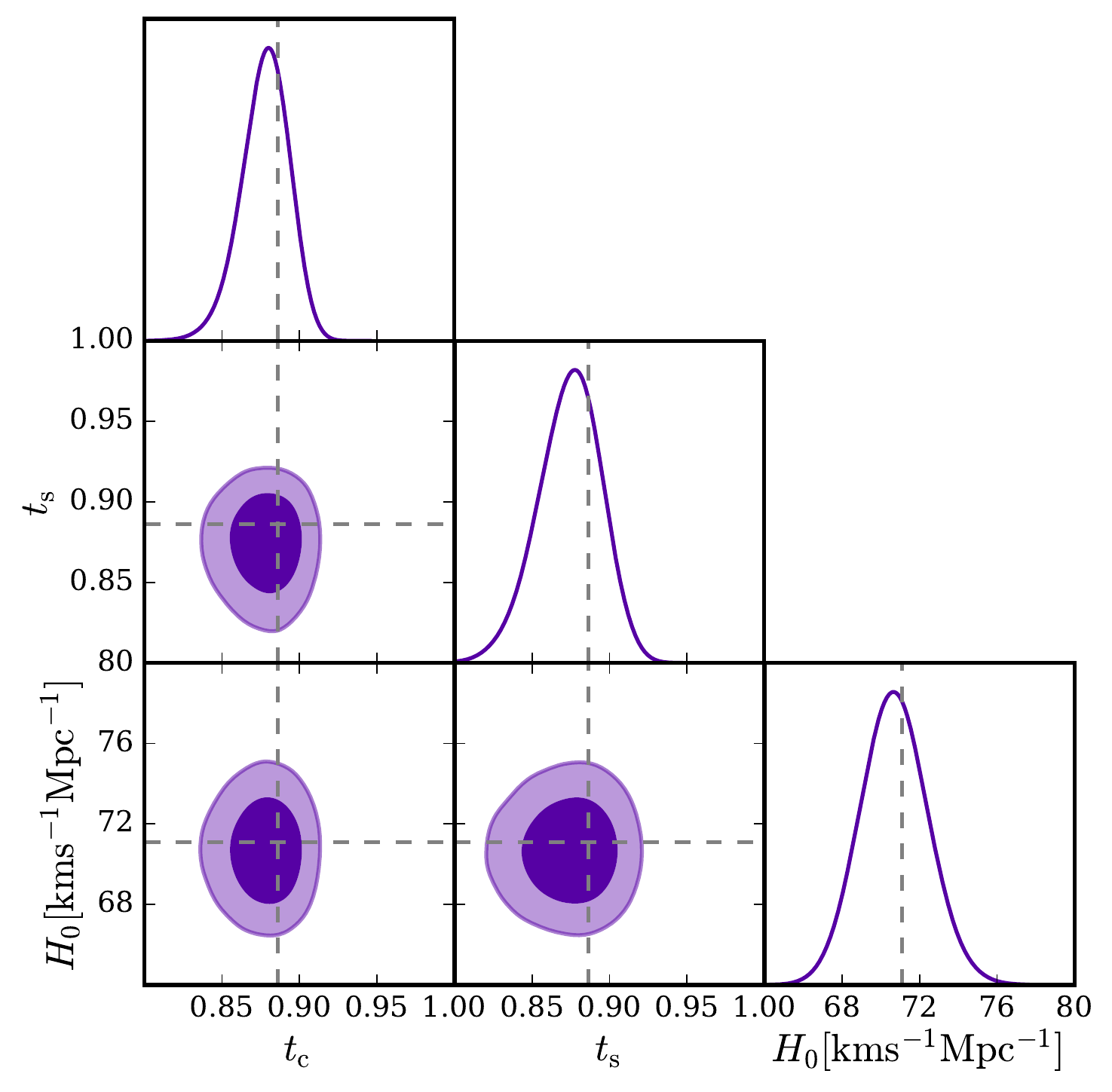}
\includegraphics[width=\figwidth]{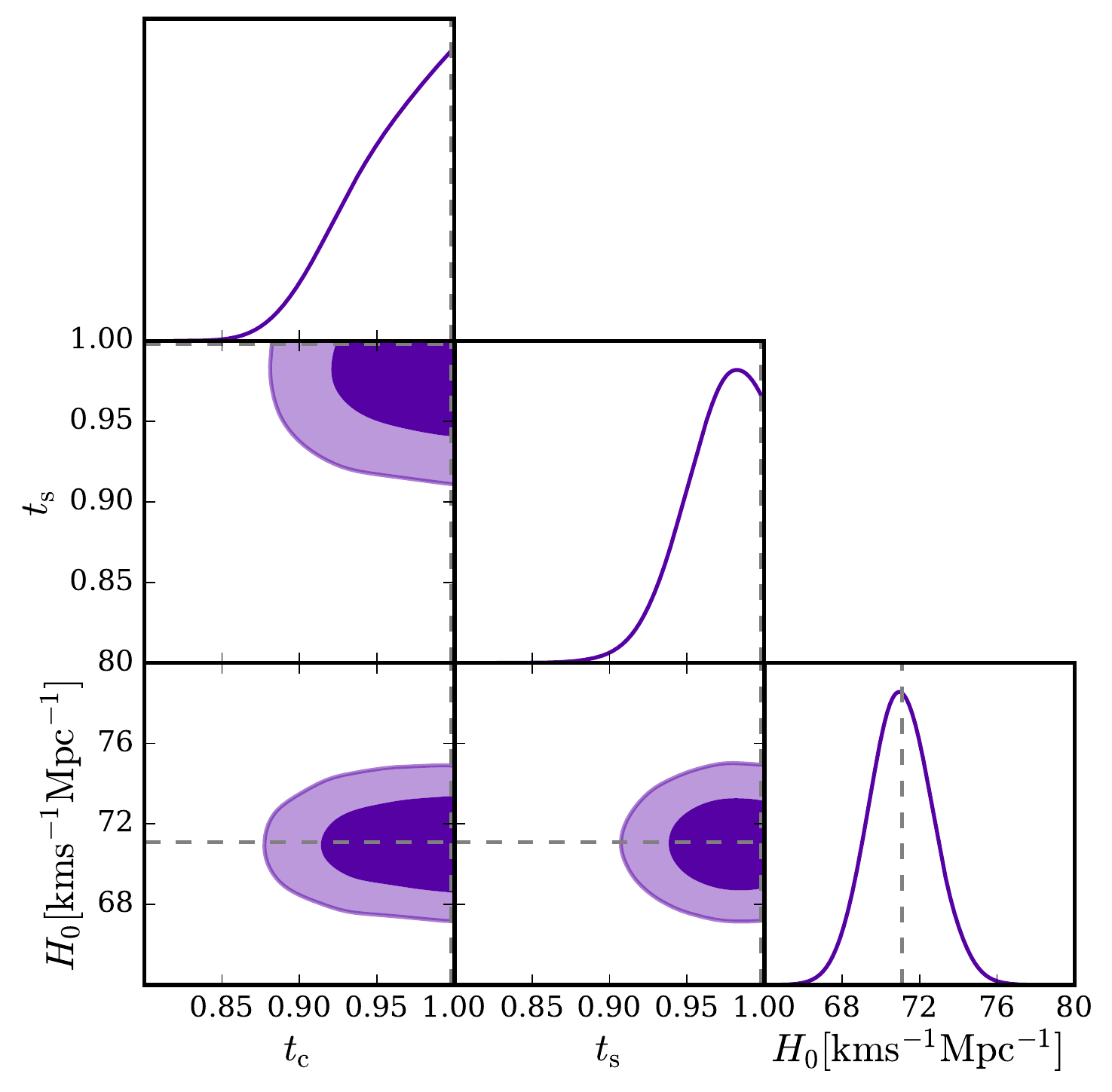}
\caption{Posteriors on $\hubble$ and kurtosis measures of Cepheid ($\tpeakc$) and SN ($\tpeaks$) intrinsic scatter distributions for three-anchor simulations with heavy-tailed (left) and Gaussian (right) scatter. The kurtosis measure, $\tpeak(\nu)$, is the ratio of the peak density of a \studentdist\ with degrees of freedom $\nu$ to that of a Gaussian with the same location and scale. Dashed grey lines indicate ground-truth parameter values.}
\label{figure:ht_triangle_sim}
\end{figure*}

In order to determine the impact of mismodeling the data -- or, more accurately, incorrectly assuming that outliers are present -- we also generate a simulation with Gaussian intrinsic scatter and infer its properties assuming the scatter is \studentdist. The results are shown in the right-hand panel of~\fig{ht_triangle_sim}. The posteriors for $\tpeakc$ and $\tpeaks$ now peak at (or very close to) one, clearly indicating the Gaussianity of the simulated data. Most importantly, the inference on $\hubble$ in this setting is unbiased, and the precision is unaffected.


\section{Analysis of observations}
\label{section:results}

\subsection{Outlier-clipped data}

We examine the three-anchor, outlier-clipped~\riess\ dataset using three variants of the BHM. In the first case, we use our vanilla version, sampling anchor likelihoods that are Gaussian in distance or parallax, respectively. The posterior on $\hubble$ derived from 200,000 total samples from the Bayesian hierarchical model is plotted as a purple solid line in the left panel of~\fig{data_h_0_comp}. Overlaid as a yellow dashed line is the log-normal posterior produced by the generalized least-squares estimator. Finally, the posteriors estimated by~\riess\ and the~\citet{Planck_XIII:2016} are overplotted as grey dashed and dot-dashed lines, respectively: we compare to the~\citet{Planck_XIII:2016} results throughout the following discussion as the MCMC samples from the \citet{Planck_Int_XLVI:2016} are not yet available. This has the effect of reducing the tension between the local and cosmological estimates to 2.8-$\sigma$: still highly significant according to its $p$-value of 0.0051.

\begin{figure*}
\includegraphics[width=0.73\figwidth]{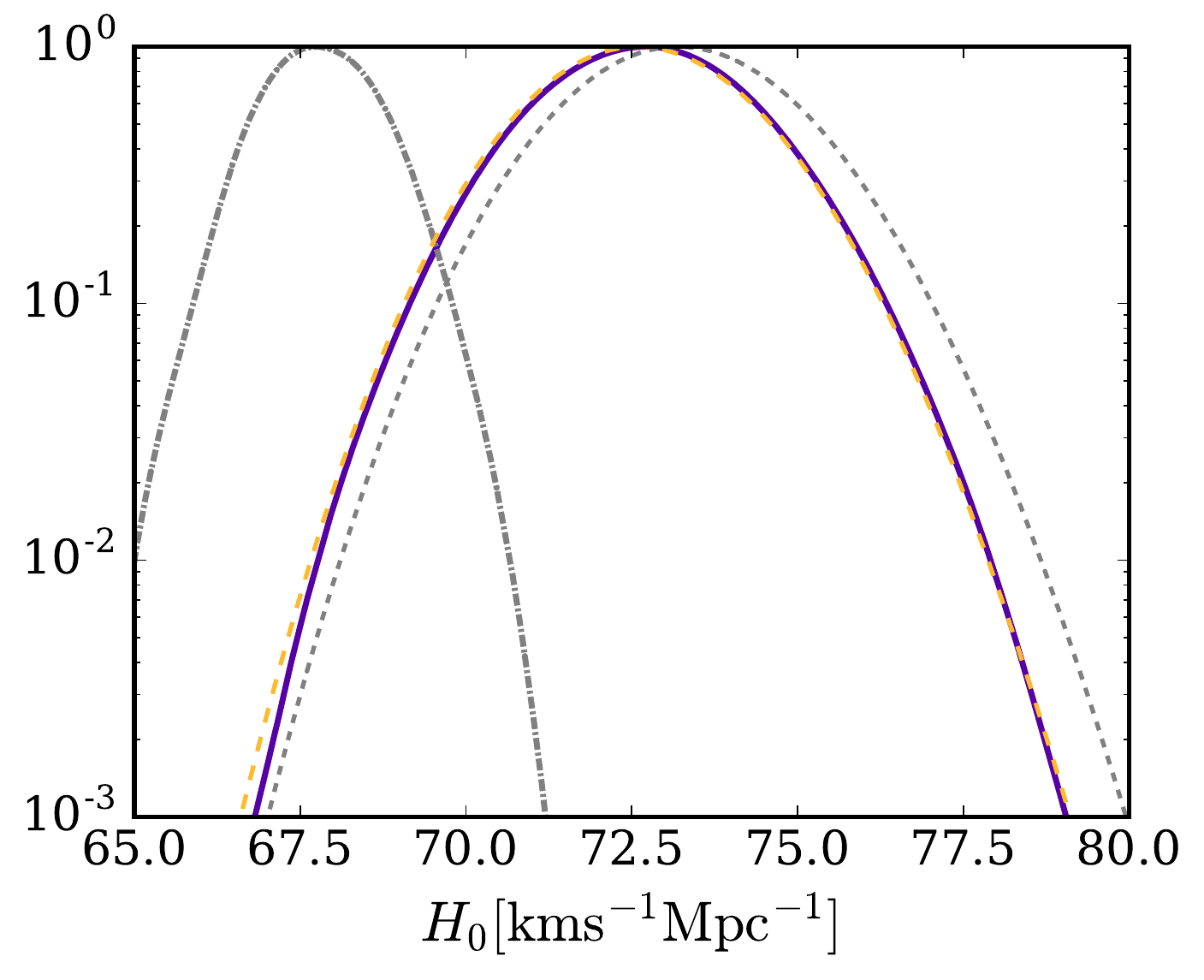}
\includegraphics[width=0.73\figwidth]{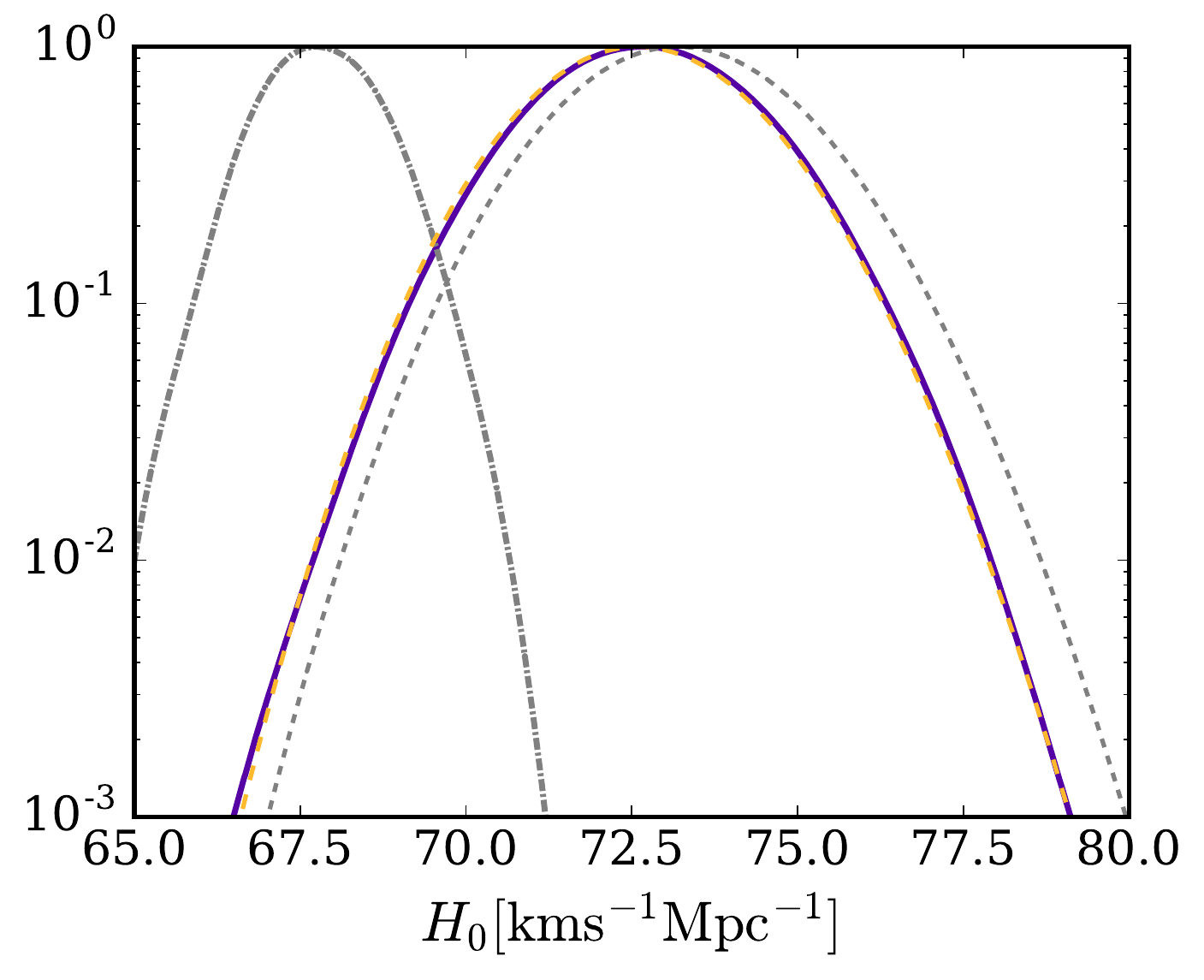}
\includegraphics[width=0.73\figwidth]{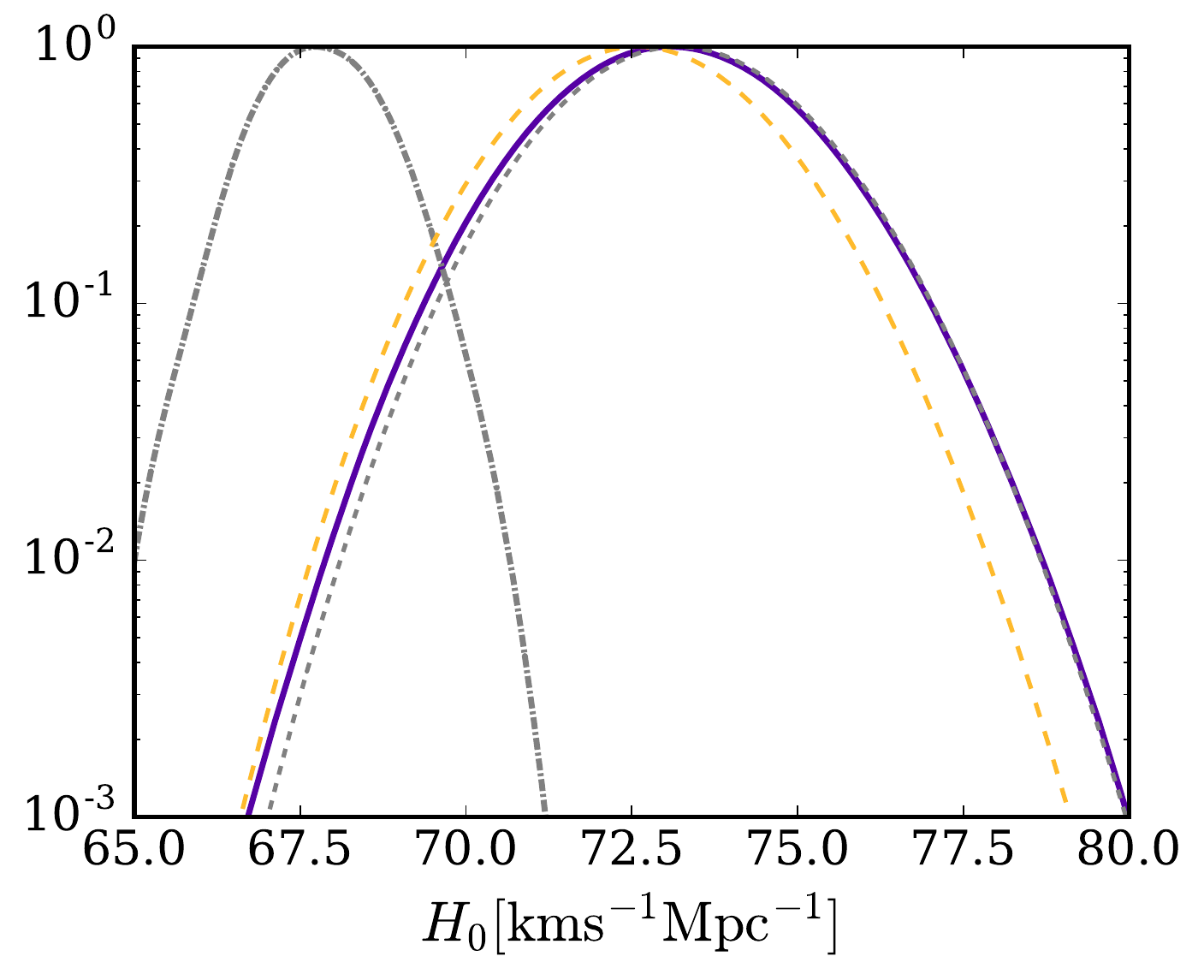}
\caption{Posteriors on $\hubble$ for the three-anchor~\protect\riess\ dataset, normalized to a peak density of one. Purple posteriors are produced by variants of the BHM: (left) sampling Gaussian anchor likelihoods (in distance/parallax) and Gaussian intrinsic scatter distributions; (centre) as previous, but substituting the non-Gaussian MASER likelihood; and (right) modeling the intrinsic scatters using \studentdist\ distributions. Log-normal posteriors from our implementation of the GLS method are plotted as yellow long-dashed lines; log-normal posteriors quoted by~\protect\riess\ are plotted as grey short-dashed lines. The posterior extrapolated from the~\citet{Planck_XIII:2016} data assuming a $\Lambda$CDM cosmology is plotted as grey dot-dashed lines.}
\label{figure:data_h_0_comp}
\end{figure*}

The estimates of $\hubble$ we obtain using both the GLS approach and the vanilla Bayesian hierarchical model are roughly 0.6 \kmsmpc\ (i.e., $\sim$1/3 of the uncertainty) lower than the fiducial value reported by \riess. Approximately half of this discrepancy (0.36 \kmsmpc) can be traced to a 0.5\% change in the treatment of the coherent velocity corrections applied to obtain the final results presented in \riess, but the remaining difference is unexplained (A. Riess, private communication). As our main aim in this paper is to explore the effects of the statistical approach on the posterior distribution in $\hubble$ we adopt our GLS estimate of (72.63 $\pm$ 1.68) \kmsmpc\ as our baseline point of comparison.

The final posterior constraint on $\hubble$ from the vanilla version of the BHM is (72.72 $\pm$ 1.67) \kmsmpc, as compared to the result from {\it Planck's} temperature, large-scale LFI polarization and lensing data of $67.81 \pm 0.92$ \kmsmpc~\citep{Planck_XIII:2016}: we find that the BHM $\hubble$ posterior drops to $1.1\times10^{-2}$ of its maximum at the central {\it Planck} value. This compares well with the density of the GLS posterior ($1.3\times10^{-2}$) but is almost twice that of the \riess\ posterior ($5.6\times10^{-3}$). As the widths of these posteriors are roughly the same, the {\it Planck} $\hubble$ value is approximately twice as likely in our BHM analysis than \riess\ would conclude, though we stress that this is driven by the discrepancy between the central values of the analyses.

In the second BHM variant, we instead model the MASER as having a non-Gaussian likelihood. Using this variant we obtain the $\hubble$ posterior as plotted in the centre panel of~\fig{data_h_0_comp}. As in simulations (\sect{outlier_sim}), using the non-Gaussian MASER likelihood pushes the posterior back toward the GLS solutions; however, the posterior constraints are stable at ($72.72 \pm 1.68$) \kmsmpc, and the BHM $\hubble$ posterior is $1.2\times10^{-2}$ of its maximum at the {\it Planck} best-fit value. This insensitivity to the functional form of the MASER likelihood is perhaps unsurprising, as in this setting we are effectively using 17 anchor likelihoods, reducing the dependence of the $\hubble$ constraint on any individual anchor.

\begin{figure*}
\includegraphics[width=\figwidth]{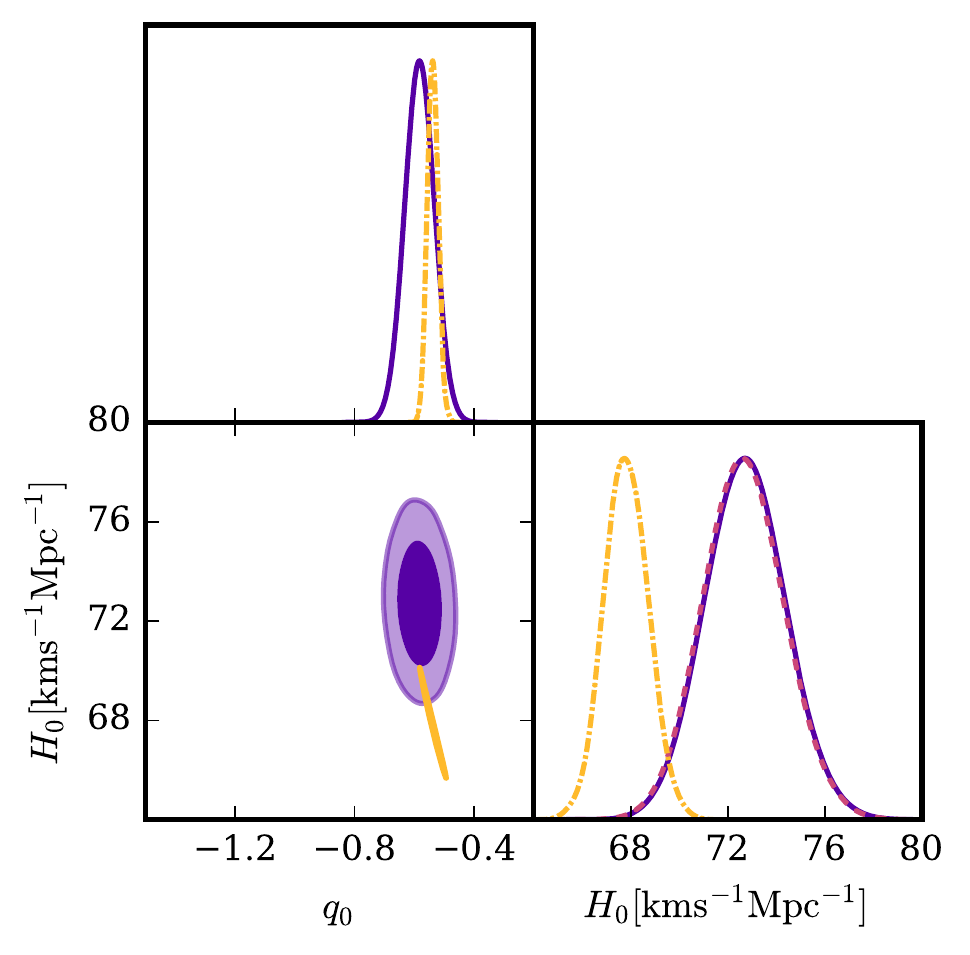}
\includegraphics[width=\figwidth]{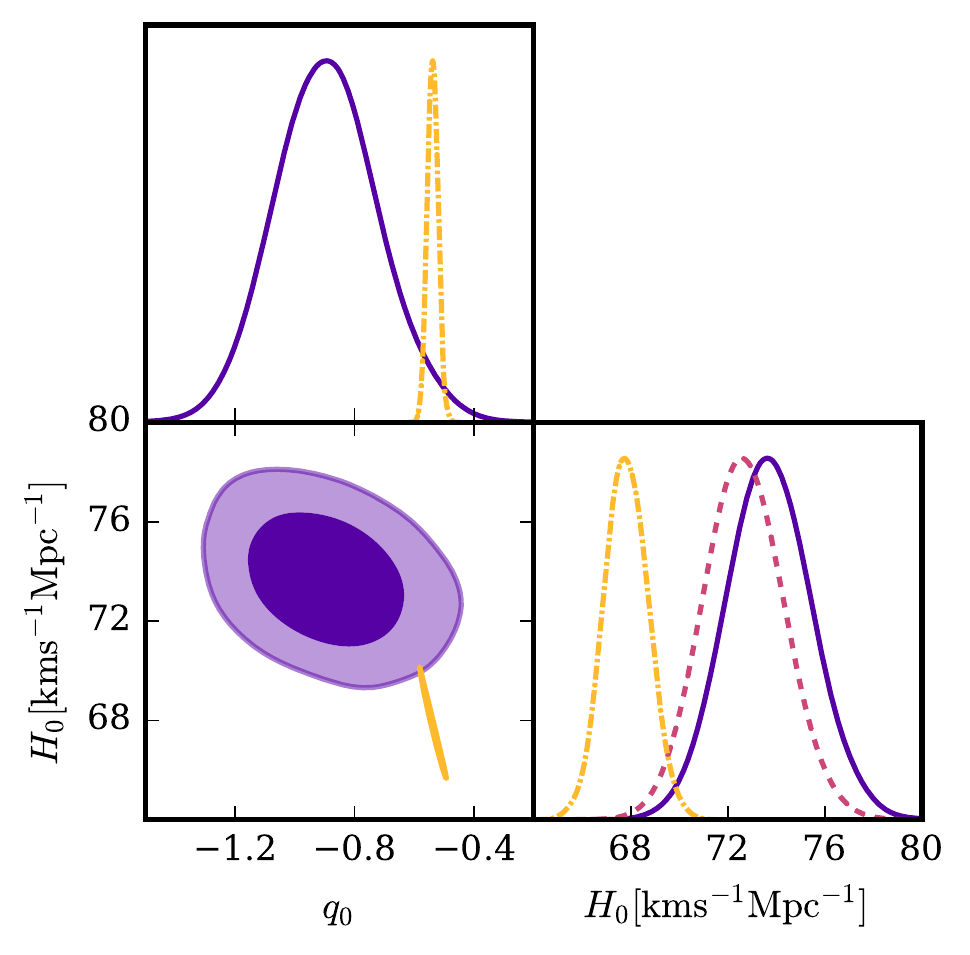}
\caption{$\hubble$ and $\decel$ posteriors derived by the BHM from the three-anchor~\protect\riess\ dataset (purple) with (left) and without (right) a tight $\Lambda$CDM-dependent $\decel$ constraint~\citep{Betoule_etal:2014}. Posteriors extrapolated from the~\citet{Planck_XIII:2016} data assuming a $\Lambda$CDM cosmology are plotted in yellow dot-dashed. Overlaid as a pink dashed line is the log-normal $\hubble$ constraint produced by the GLS fit from the local distance ladder.}
\label{figure:gauss_q_0_dep}
\end{figure*}

It is also instructive to examine the dependence of our $\hubble$ constraints on the deceleration parameter, $\decel$. In the left panel of~\fig{gauss_q_0_dep} we show the joint and marginal posteriors on $\hubble$ and $\decel$ estimated by our vanilla BHM including the $\Lambda$CDM-specific~\citet{Betoule_etal:2014} constraint on $\decel$ (as employed by \riess); the~\citet{Planck_XIII:2016} posteriors are overlaid in yellow. In the right panel, we remove the~\citet{Betoule_etal:2014} constraint, treating $\decel$ as a nuisance parameter constrained weakly by the distance ladder data and its broad (truncated Gaussian) prior. The degeneracy this opens up does not affect the precision of the $\hubble$ constraint significantly, but the mean-posterior value increases by 1 \kmsmpc to ($73.70 \pm 1.77$) \kmsmpc. While removing the observational $\decel$ constraint from the analysis may seem somewhat extreme, its cosmological model dependence, coupled with the fact it is derived using at least some of the same data as used in \riess, calls into question its general applicability. This strongly motivates extending the current BHM to include the~\citet{Betoule_etal:2014} dataset to better constrain $\decel$ in a model-independent fashion, although this is left for future work. In the following, we remove the~\citet{Betoule_etal:2014} constraint in order to make conservative statements, independent of any cosmological model.

Having established the insensitivity of the parameter constraints to the form of the MASER likelihood, we turn towards the question of model selection.  We are now able to quantify the discrepancy between local measures of the expansion and their expectations from the CMB using model selection. As the distance ladder data are sensitive to, and {\it Planck} makes firm predictions for, both the expansion rate and deceleration, we must include both parameters in our model selection analysis. We therefore compare two models: one in which the cosmological and local $\hubble$ and $\decel$ are identical (denoted $\msame$ following the notation of \sect{stats}), and a second (denoted $\mdiff$) in which the cosmological values differ by $\dhubble$ and $\ddecel$. Note that the simple model is nested within the complex model at $\dhubble = \ddecel = 0$: this is a simple extension to the one-additional-dimensional example of \sect{stats}. As the models are nested, we can simply adapt the BHM (as shown in~\fig{network_mod_sel}) to estimate the joint posterior of $\dhubble$ and $\ddecel$, then compare the posterior to the prior at $\dhubble = \ddecel = 0$ to obtain the Bayes factor for the two models via the Savage-Dickey Density Ratio~(SDDR; \citealt{Dickey:1971}), which gives the Bayes factor between the models in terms of a posterior density ratio as
\begin{equation}
\left. \frac{ \prob(\datl , \datc | \msame) }{ \prob(\datl , \datc |  | \mdiff) } = \frac{ \prob(\dhubble , \ddecel | \datl , \datc , \mdiff) }{ \prob(\dhubble |\mdiff) \, \prob(\ddecel | \mdiff) } \right|_{\dhubble = \ddecel = 0}.
\end{equation}

We exploit two {\it Planck} datasets in our model-selection analysis. We take the less-discrepant but complete~\citet{Planck_XIII:2016} release as our standard, as its MCMC chains are available for download. Using these chains, we summarize the~\citet{Planck_XIII:2016} joint posterior on $\hubble$ and $\decel$ as a strongly correlated bivariate Gaussian, centred on (67.81 \kmsmpc, $-0.5381$), with standard deviations of (0.92 \kmsmpc, $0.0184$) and a correlation coefficient of $-0.99$. Ideally, we would repeat the analysis using the more-recent, and more-discrepant, \citet{Planck_Int_XLVI:2016} intermediate release; however, these MCMC chains have not been made public. Instead, we approximate the~\citet{Planck_Int_XLVI:2016} posterior by applying a $\normal(\tau; 0.055, 0.009^2)$ prior to the public~\citet{Planck_XIII:2016} likelihood (A. Riess, private communication). Combining this $\tau$ prior with the ``TT,TE,EE+lowP'' likelihood and sampling using CosmoMC~\citep{Lewis:2002ah,Lewis:2013hha}, we obtain an $\hubble$-$\decel$ posterior well approximated by a bivariate Gaussian with mean (66.74 \kmsmpc, -0.5155), standard deviations (0.62 \kmsmpc, 0.0132) and correlation coefficient -0.994.\footnote{The means and standard deviations from the corresponding~\citet{Planck_Int_XLVI:2016} ``TTTEEE+SIMlow'' analysis are $\hubble = (66.93 \pm 0.62)$ \kmsmpc\ and $\decel = -0.5197 \pm 0.0131$. Our approximated {\it Planck} $\hubble$ constraint is slightly {\it more} discrepant from \riess\ than the~\citet{Planck_Int_XLVI:2016} estimate, and the resulting odds ratios are therefore slightly more harsh on $\Lambda$CDM than those we would obtain using the true posteriors.} While the conclusions we derive using this second {\it Planck} dataset are necessarily approximate, they provide a useful guide to the impact of the latest {\it Planck} observations.

\begin{figure}
\includegraphics[width=\columnwidth]{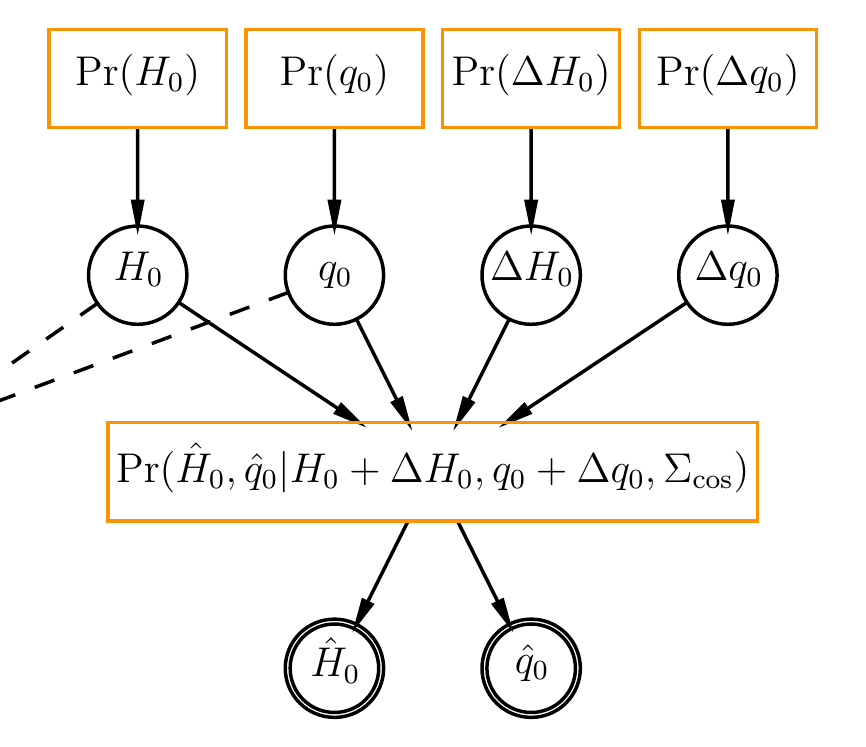}
\caption{Extension to Bayesian hierarchical model to allow selection between models in which the local and cosmological Hubble ($\hubble$) and deceleration ($\decel$) parameters are the same or differ by $\dhubble$ and $\ddecel$, respectively. Dashed lines indicate connections to the Hubble flow SN portion of the hierarchical model used for inference of the local data only (\fig{network}).}
\label{figure:network_mod_sel}
\end{figure}

Bayesian model comparison results are strongly dependent on the adopted parameter priors, which is a potential concern when considering phenomenological models such as the more complex model $\bar{\Lambda}$ above. Following \sect{stats}, we select these priors to construct a lower bound on the probability of the simple model ($\msame$) subject to the chosen form of the extended model ($\mdiff$). We restrict ourselves to considering models predicting symmetric (Gaussian) deviations away from the true values of $\hubble$ and $\decel$. To construct our lower bound, we first set the prior probabilities of the two models to 0.5, the harshest penalty one should apply to a concordance model. Second, we choose the standard deviation of the Gaussian priors on $\dhubble$ and $\ddecel$ to be 6.0 \kmsmpc\ and 0.5, respectively, creating a model designed to explain the discrepancies in the data. We use the same standard deviations for the Gaussian priors on the true underlying values of $\hubble$ and $\decel$: the results should be independent of the precise choice for priors this broad; we centre the priors on 70 \kmsmpc\ and $-0.7$.

Applying the SDDR to the outputs of our BHM given the above prior choices, we find the Bayes factor to be $0.11 \pm 0.01$ using the standard~\citet{Planck_XIII:2016} likelihood. We would therefore state that (assuming equal {\it a priori} model probabilities) an extended model of the form considered here is {\em at most} nine times more likely than $\Lambda$CDM to be the true model, given the CMB and distance-ladder data. Swapping in our approximation to the more-discrepant~\citet{Planck_Int_XLVI:2016} likelihood, this odds ratio decreases significantly, yielding an odds ratio of $0.016 \pm 0.005$ or roughly 1:60. While still strongly in favour of the extended model, these probability ratios are significantly less extreme than the corresponding $p$-values would suggest, and, most critically, have a clear interpretation. Furthermore, the framework is simple to apply to constraints from physically motivated models. 

\subsection{Analysis including SN outliers}

In our final analysis, we reintroduce the SN outliers to the high-redshift dataset and use the heavy-tailed intrinsic scatter distributions in our hierarchical model. While the~\riess\ dataset containing Cepheid outliers is not available, we can still model the Cepheid intrinsic scatter as heavy-tailed as a test of both data and model. Given the results derived from similar simulations (\sect{outlier_sim}), we expect in this case that the posterior would peak at $\tpeakc = 1$, corresponding to the Gaussian limit that $\dofc \rightarrow \infty$.

\begin{figure*}
\includegraphics[width=1.1\figwidth]{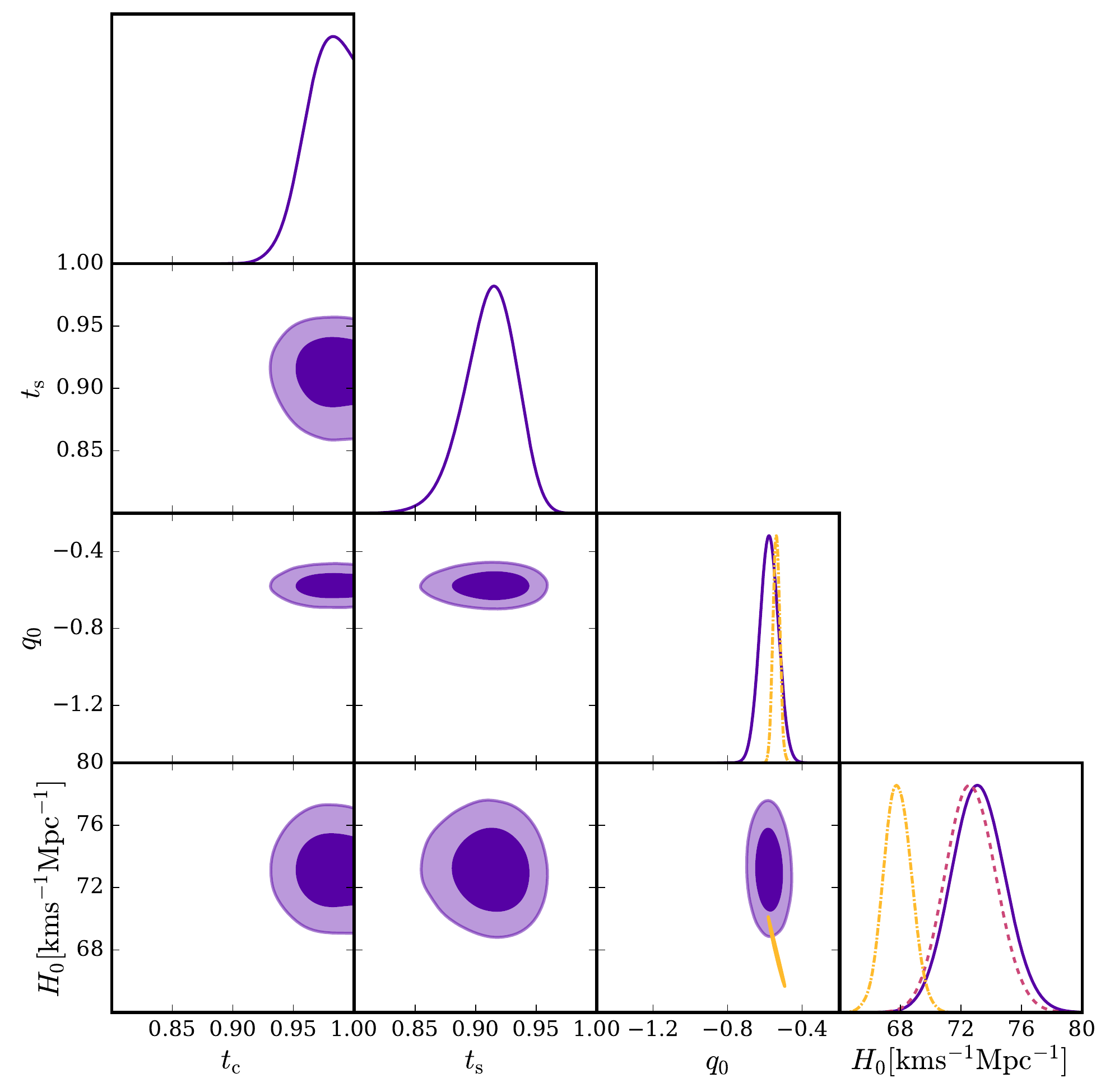}
\includegraphics[width=1.1\figwidth]{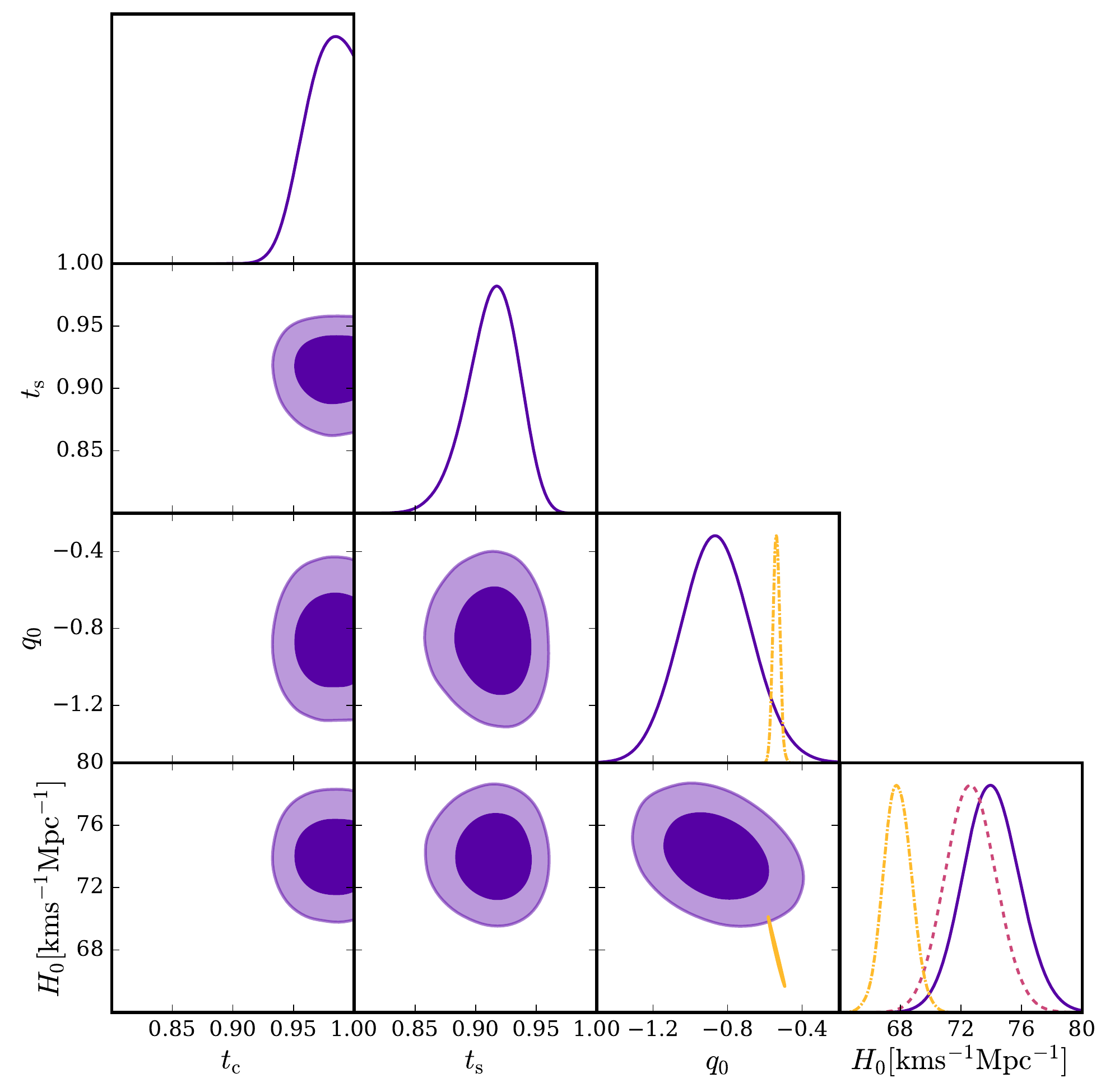}
\caption{Posteriors on $\hubble$, $\decel$ and the shapes of the Cepheid ($t_{\rm c}$) and SN ($t_{\rm s}$) intrinsic scatter distributions (purple) for the three-anchor~\protect\riess\ dataset including SN outliers, with (left) and without (right) a tight $\Lambda$CDM-dependent $\decel$ constraint~\citep{Betoule_etal:2014}. Taken to be \studentdist\ distributions, the intrinsic scatters are Gaussian in the limit $t \rightarrow 1$. The log-normal $\hubble$ constraint derived from the GLS fit is overlaid as a pink dashed line. Posteriors extrapolated from~\citet{Planck_XIII:2016} data assuming a $\Lambda$CDM cosmology are plotted in yellow dot-dashed.}
\label{figure:ht_triangle}
\end{figure*}

This is indeed the case, as demonstrated in the marginalized posteriors shown inthe left panel of \fig{ht_triangle}. The posterior of $\tpeakc$ peaks at a value of 0.98, with a 95\% posterior lower limit of 0.94: the Cepheid scatter distribution is very close to Gaussian. This is especially clear when compared to the SNe, for which $\tpeaks = 1$ is strongly disfavoured (we find a $\dofs = 2.9 \pm 0.8 $). Turning to the $\hubble$ posterior (see also \fig{data_h_0_comp}, right panel), we see a small but distinct shift to higher values of $\hubble$ when modeling outliers instead of removing them, more in line with the findings of~\riess\ and~\citet{Cardona_etal:2016}: the posterior is ($73.15 \pm 1.78$) \kmsmpc. The posterior is also slightly broader, with slower-than-Gaussian decay in the tails, but the {\it Planck} best-fit $\hubble$ value is still less likely than previously found: the BHM $\hubble$ posterior is $8.9\times10^{-3}$ of its maximum at the {\it Planck} $\hubble$.

The impacts of removing the~\citet{Betoule_etal:2014} $\decel$ constraint can be seen in the right panel of \fig{ht_triangle}. The intrinsic scatter distributions inferred from the data do not change, but the $\hubble$ posterior shifts to even higher values, with ($74.02 \pm 1.86$) \kmsmpc. In terms of model selection, using the SDDR we find that the Bayes factor between $\Lambda$CDM and the extended model is now $0.15 \pm 0.01$ using the standard~\citet{Planck_XIII:2016} likelihood, or $0.016 \pm 0.005$ using our approximation to the~\citet{Planck_Int_XLVI:2016} likelihood. If the two models are equally likely {\it a priori}, the extended model is no more than seven or sixty times more likely than $\Lambda$CDM to describe the CMB and distance-ladder data, including outliers, depending on whether the standard or latest {\it Planck} data are employed.


\section{Conclusions}
\label{section:conc}

We have used Bayesian inference to assess the apparent discrepancy between the value of the Hubble constant obtained from local distance ladder measurements and that implied by the {\em Planck} CMB data.  The correct interpretation of the apparently significant 2.8 to 3.4-$\sigma$ discrepancy requires a model comparison calculation, from which it is clear that the critical quantity is not simply the difference between the two quoted $\hubble$ estimates: the result depends on the tails of the likelihoods from the two data-sets. In this particular situation the fact that the cosmological estimate of $\hubble$ has a smaller uncertainty means that it is the low-$\hubble$ tail of the local distance ladder likelihood that is critical.  Taking a model-comparison approach does not penalize the null model (that there is no discrepancy) as strongly as, \eg, using overly-punitive $p$-values; but to make any quantitative statements requires the full likelihood from the local distance ladder, as opposed to a Gaussian or least squares approximation.

We have hence reframed the local distance ladder as a BHM, in which all components are simultaneously and self-consistently inferred.  By using HMC, as implemented in Stan, we are not restricted to Gaussian/normal assumptions, allowing us to use heavy-tailed distributions (specifically the student-t distribution) to obtain robust inferences even in the presence of outliers.  Using simulations, we analyze the performance of the model and sampler in a controlled setting, compare with existing generalized least squares techniques, and investigate extensions to the model to account for outliers in the Cepheid and SN populations.
We have avoided rerunning the analysis with different models, anchors, redshift ranges, \etc\ (\cf\ \riess\ and the blind analysis of \citealt{Zhang_etal:2017}), to gauge any systematic effects, as by modeling the outlier population we have attempted to remove data cuts from the analysis. Most of these other choices could be coded into a hierarchical model (although the priors on data selection would be challenging to formulate), something which is left to future work.

When applied to distance ladder data that have been cleaned of outliers (using the sigma-clipping techniques of \riess), we find a slight ($-0.6$ \kmsmpc) discrepancy between our posterior constraint of $\hubble$ = (72.72 $\pm$ 1.67) \kmsmpc\ and that of \riess: this is not present in comparisons between the two algorithms in simulations. Our $\hubble$ posterior drops to $1.1\times10^{-2}$ of its maximum at {\it Planck's} most likely value of 67.81 \kmsmpc. When we reintroduce SN outliers to the dataset (the Cepheid outliers are not available) and allow both Cepheids and SNe to have heavy-tailed scatter about their Leavitt and Tripp laws, respectively, the posterior constraint on $\hubble$ shifts upwards to ($73.15 \pm 1.78$) \kmsmpc, which is much closer to that of \riess. Although the resulting posterior is also slightly broader, it drops to $8.9\times10^{-3}$ of its maximum at {\it Planck's} central $\hubble$ value.  Variations in SN outlier mitigation can, therefore, affect our inference on $\hubble$ non-trivially, but not enough to explain the discrepancy with the {\it Planck} data. The tension between the local estimate of $\hubble$ and the value extrapolated from {\it Planck} assuming a flat $\Lambda$CDM cosmology persists when the tails of the local $\hubble$ posterior are accurately characterized.

We also note the dependence of this tension on the treatment of the deceleration parameter, $\decel$, a nuisance parameter that is typically fixed when inferring $\hubble$. The above results allow $\decel$ to vary, but only within observational bounds derived assuming a flat $\Lambda$CDM cosmology that in turn makes use of an overlapping SN dataset. Removing this observational constraint opens a degeneracy with $\hubble$, shifting the mean posterior values up to ($73.70 \pm 1.77$) \kmsmpc, or ($74.02 \pm 1.86$) \kmsmpc\ if the (SN) outliers are included and fit with heavy-tailed intrinsic scatter distributions.

We quantify the tension between the local and cosmological expansion measurements in a principled fashion using Bayesian model comparison, with parameter priors chosen to minimize the posterior of the null model, $\Lambda$ (\ie, that $\Lambda$CDM holds). We compare against a designer model, $\mdiff$ in which the cosmological $\hubble$ and $\decel$ are allowed to differ from their local counterparts on a scale set by the observed discrepancies. We find Bayes factors of $0.11 \pm 0.01$ and $0.15 \pm 0.01$ between these models when using the standard (less-discrepant)~\citet{Planck_XIII:2016} likelihood, depending on whether SNe outliers are cleaned or modeled. This indicates that the designer model is {\em at most} nine times more likely than $\Lambda$CDM to be the true model, given the CMB and outlier-free distance-ladder data, and no more than seven times more likely than $\Lambda$CDM when SNe outliers are included and modeled. While the more-discrepant~\citet{Planck_Int_XLVI:2016} likelihood is not publicly available, we are able to approximate it using a tight prior on $\tau$. Doing so reduces the Bayes factors for both outlier-mitigation techniques considerably to $0.016 \pm 0.005$, implying that the designer model is at most sixty times more likely than $\Lambda$CDM when the latest {\it Planck} data are considered.

An important feature of the hierarchical model presented here is that it can easily be made more realistic or extended to include more datasets. The most obvious next step would be to include the colour and light-curve shape of the Cepheid-host SNe (as in~\cite{Zhang_etal:2017}), and hence fully couple the nearby and Hubble flow SNe. Replacing the generic heavy-tailed likelihood with physical models of the Cepheid and SN outliers would improve the model's fidelity and make better use of existing data. Extending the redshift range of the model would allow simultaneous constraints of $\hubble$ and the deceleration parameter, and modifying the Tripp relation would probe the robustness of the resulting conclusions.  More locally, the Cepheid portion would be reinforced by allowing for errors in the periods and metallicities.  While the model was designed explicitly for the task of analyzing the local distance ladder, it could easily be adapted to a variety of structurally similar problems in astronomy.  A particular promising avenue would be to use it to link populations at different distances in the {\em Gaia} data, for which similar models have already been deployed (\eg, \citealt{Sesar_etal:2016}).  


\section*{Acknowledgments}

SMF thanks Justin Alsing, Lauren Anderson, David Hogg, Andrew Jaffe, Leon Lucy, Adam Riess, Dan Scolnic and David Spergel for useful discussions, and Chloe Calder for inspiration. The Flatiron Institute is supported by the Simons Foundation. SMF was partially funded by STFC in the United Kingdom. The authors acknowledge the use of the Daft\footnote{\url{http://daft-pgm.org/}} and GetDist\footnote{\url{http://getdist.readthedocs.io/}} Python packages.


\bibliographystyle{mnras}
\bibliography{references}


\appendix
\section{Student-t distribution as a model for outliers}
\label{section:dof_prior}

One of the fundamental principles of Bayesian inference,
and of this calculation in particular, 
is that the data under consideration be predicted by a model,
rather than processed or modified in any way.
One potential difficulty in implementing this principle is that
real datasets contain outliers:
a small fraction of measurements that are very different from the 
main population, and hence anomalous.
These are typically described as cases where the measurement process has 
gone wrong, although it is really more correct to say that the 
model is inadequate, being incapable of predicting something that 
actually occurs with sufficient frequency to be observed even in a
typical dataset.

It is also possible for the outliers to be independent of any 
measurement, as is the case here for some SNe and,
in particular, for the small number of Cepheids which lie away from the main 
population in period-luminosity(-metallicity) space.
The treatment of these objects in efforts to measure 
$\hubble$ has was sufficiently prominent in 
\cite{Riess_etal:2011} that this was the main focus of the
reanalysis by \cite{Efstathiou:2014};
and one of the main changes described by \riess\ is 
the much lower fraction of outliers removed from the Cepheid data.

Ideally, however, there would be no outlier removal, and instead
the adopted model would be able to generate fully realistic data,
including outliers at the appropriate frequency and 
with the observed properties.
This typically means replacing normal distributions 
(often chosen because of their numerical convenience,
although potentially motivated by the central limit theorem
or maximum entropy considerations) 
with distributions that have heavier tails.
The inevitable result is that the inference calculation becomes 
numerically more complicated, 
to a degree which often outweighs the conceptual simplicity of 
having (or at least aiming to have) a model which can generate the 
full dataset.

A flexible choice for a heavy-tailed distribution
is the Student-t distribution, defined 
in scaled and translated form by the density 
\begin{equation}
\label{equation:student}
\student(x; \mu, \sigma, \nu)
  = \frac{\Gamma\left(\frac{\nu + 1}{2}\right)}
    {(\pi \, \nu)^{1/2} \, \Gamma\left(\frac{\nu}{2}\right) \, \sigma}
    \left[1 + \frac{(x - \mu)^2}{\nu \, \sigma^2} \right]^{-(\nu + 1)/2},
\end{equation}
where 
$\Gamma(t) = \int_0^\infty \diff x \, x^{t - 1} \, e^{-x}$ is the 
Gamma function,
the peak of the distribution is at $x = \mu$,
the width of the distribution scales with $\sigma \geq 0$,
and $\nu > 0$ controls the shape of the distribution,
with heavier tails for lower values of $\nu$.
The wings of the Student distribution are sufficiently broad that its
$n^{\rm th}$ moment is only defined if $\nu > n$, so that
for $\nu \leq 2$ the Student distribution has no formal variance,
and for $\nu \leq 1$ it does not have a mean either.
Most importantly,
$\lim_{\nu \rightarrow \infty} 
\student(x; \mu, \sigma, \nu) = \normal(x ; \mu, \sigma^2)$,
where 
\begin{equation}
\label{equation:normal}
\normal(x ; \mu, \sigma^2)
  =
  \frac{1}{(2 \pi)^{1/2} \, \sigma}
  \exp \left[ -\frac{1}{2}
  \left( \frac{x - \mu}{\sigma} \right)^2
  \right],
\end{equation}
which is a Gaussian distribution of mean $\mu$ and variance $\sigma^2$.
This means that if the Student distribution is used to model a dataset
for which there are no outliers the possibility of a normal model
is retained.

The values of $\mu$ and $\sigma$ are linked to the core of the 
population being modeled; it is the value of $\nu$ which 
determines the prevalence of outliers. 
In general this is not likely to be known 
(or derivable) {\it a priori}, 
implying that $\nu$ 
is a parameter to be inferred from the data.
An immediate corrollary of this is that a prior distribution,
$\prob(\nu)$, for 
$\nu$ is required.
While there is no compelling form for this prior, 
it is clear that 
the Gaussian ($\nu \rightarrow \infty$) limit 
should neither be precluded
(in which case heavy tails would be enforced even if the data did
  not demand it)
nor strongly favoured
(as would be the case if an improper uniform for $\nu$ were adopted).
It is also useful that very heavy-tailed distributions,
with $\nu < 1$ be included, although there is an expectation that such 
models, which are effectively dominated by outliers,
would be ruled out by most useful datasets.

A number of possible priors have been explored with the aim of satisfying at least some of these (or similar) requirements:

\begin{itemize}

\item
A sensible starting point is to choose a prior on $\nu$ such that the kurtosis of the resulting \studentdist\ distributions is uniformly sampled; however, the kurtosis of the standard \studentdist\ distribution is $6/(\nu-4)$, which is only defined for $\nu > 4$. The implied (Pareto) prior, $\prob(\nu) = \step(\nu - \nu_{\rm min}) / (\nu-4)^2$ requires $\nu_{\rm min} > 4$, omitting a potentially important region of parameter space. \cite{Rubio_Steel:2014} suggest bypassing this issue by using a different kurtosis measure (based on the ratio of the mode of the distribution to its lower inflection point). The resulting prior has a somewhat involved functional form, but the concept of using a prior uniform in some heuristic for the kurtosis is a promising avenue.

\item
\cite{Juarez_Steel:2010} adopted a Gamma distribution with shape 2 and rate 0.1, with $\prob(\nu) \propto \step(\nu) \, \nu^2 \, e^{-\nu / 10}$, which has since been advocated for by the Stan community.  However, this choice of prior results in a somewhat arbitrary focus on $\nu \simeq 10$ which is not justified by any obvious underlying argument.

\item
\cite{Ding:2014} used a Gamma distribution with shape 1 and rate 0.1, which reduces to an exponential distribution with $\prob(\nu) \propto \step(\nu) \, e^{-0.1 \,\nu}$. While this prior produces 
more low-$\nu$ draws than the \cite{Juarez_Steel:2010} prior, the exponential suppression for $\nu$ much larger than the inverse rate renders the Gaussian limit unattainable in practice.

\end{itemize}

Motivated in part by the shortcomings of the above priors,
we advocate a different option here,
which is to choose a prior on $\nu$ such that the 
resultant distribution of peak heights is uniform: the peak height is, 
in essence, a heuristic for the kurtosis of the sampled distributions.
This allows the full range of shapes, including the Gaussian limit, 
but included in such a way that it is not strongly favoured.
The peak density of the \studentdist\ distribution,
relative to the Gaussian limit, is 
\begin{equation}
\label{equation:tpeak}
\tpeak = \frac{T(x; \mu, \sigma, \nu)}{N(x; \mu, \sigma^2)} =  
\frac{2^{1/2} \, \Gamma\left(\frac{\nu + 1}{2}\right)}{\nu^{1/2} \, \Gamma\left(\frac{\nu}{2}\right)}
= \begin{cases}
\left( \frac{\pi}{2} \, \nu \right)^{1/2} & \text{if } \nu \ll 1;\\
1 - \frac{1}{4 \, \nu} & \text{if } \nu \gg 1.
  \end{cases}
\end{equation}
We plot the relationship between $\nu$ and $\tpeak$ in the left panel of~\fig{dof_prior}.

\begin{figure*}
\includegraphics[width=\figwidth]{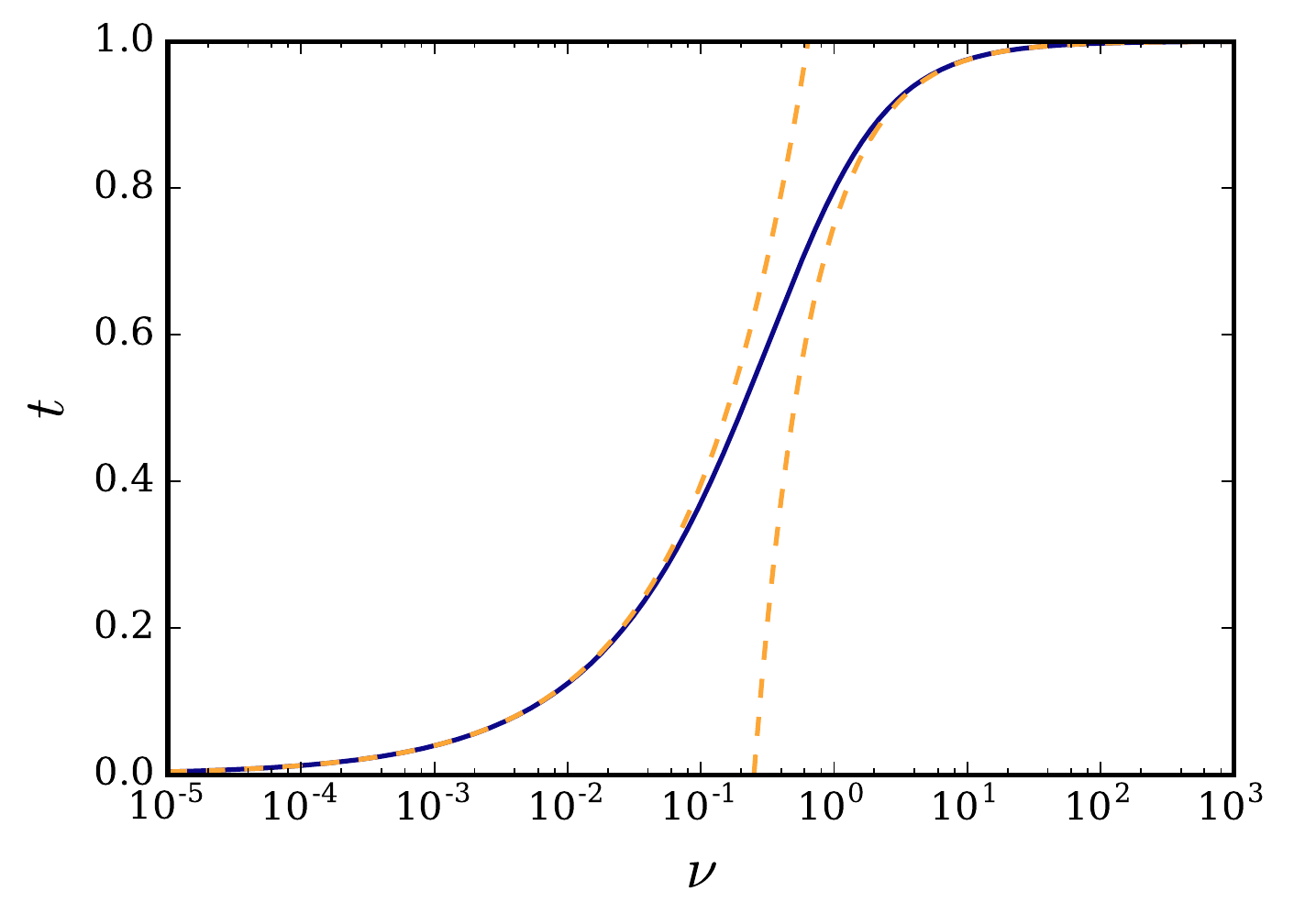}
\includegraphics[width=\figwidth]{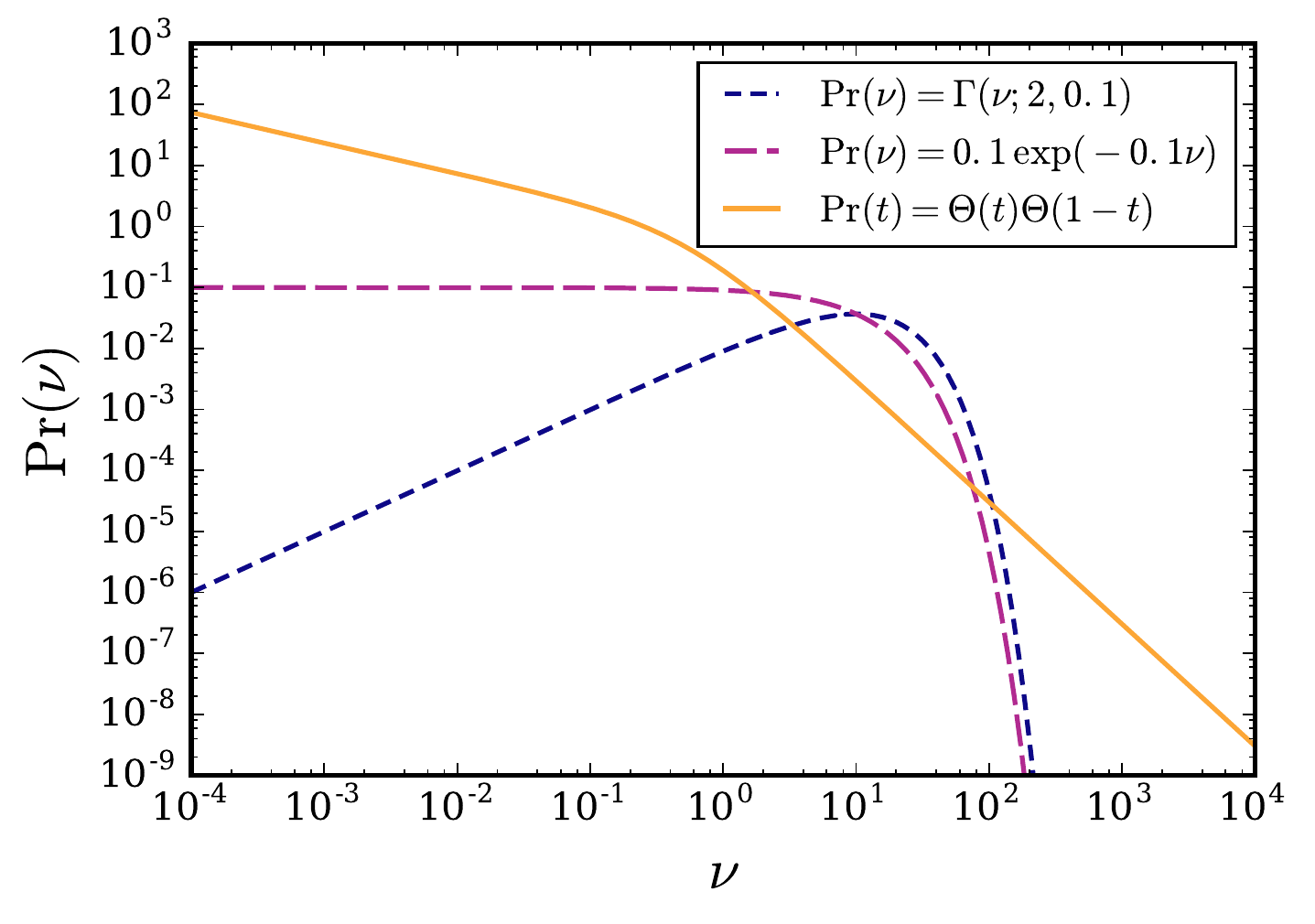}
\caption{Left: relationship between our \studentdist\ kurtosis heuristic -- the peak density of the \studentdist\ distribution relative to the Gaussian limit -- and the distribution's degrees of freedom. The limiting cases described in~\eq{final_dof_prior} are highlighted in dashed orange. Right: priors considered for the degrees of freedom of the heavy-tailed Cepheid and SN intrinsic scatter distributions.}
\label{figure:dof_prior}
\end{figure*}

The prior distribution in $\nu$ that produces a uniform distribution
of $\tpeak$ between 0 and 1 is then given by applying a Jacobian
transformation to obtain
$\prob(\nu) \propto \step(\nu) \, |\diff \tpeak / \diff \nu|$.
Unfortunately, 
the gamma function does not have a simple
derivative, so this distribution cannot be written down, 
although \eq{tpeak} implies that the 
asymptotic forms of the prior should be
$\prob(\nu) \propto \nu^{-1/2}$ if $\nu \ll 1$
and $\prob(\nu) \propto \nu^{-2}$ if $\nu \gg 1$ 
(which matches the Pareto prior discussed above).
A simple, continuous, density that 
satisfies these requirements is 
\begin{equation}
\prob(\nu) \propto 
  \frac{\step(\nu)}{\left[(\nu / \nu_0)^{1 / (2 \, a)} 
    + (\nu / \nu_0)^{2 / a}\right]^a},
\label{equation:final_dof_prior}
\end{equation}
where taking $\nu_0 \simeq 0.55$ and $a \simeq 1.2$
results in a 
peak height distribution that is within a few percent of being uniform.
We plot this prior along with the gamma and exponential distributions of 
\cite{Juarez_Steel:2010} and \cite{Ding:2014} in the right panel of~\fig{dof_prior}.
In~\fig{dof_prior_quantiles}, we plot \studentdist\ distributions corresponding 
to quantiles of the three priors, indicating how well each prior samples the 
range of possible heavy-tailed distributions. Solid curves are plotted for the  
10$^{\rm th}$, 20$^{\rm th}$, $\ldots$, 80$^{\rm th}$, and 90$^{\rm th}$ percentiles, 
and dashed lines indicate the 0.1$^{\rm st}$ and 99.9$^{\rm th}$ percentiles; the 
curves are coloured by increasing $\nu$. The tendency of the gamma and exponential 
priors to favour near-Gaussian distributions is clear, as is the uniform sampling of 
peak heights achieved by the prior described by~\eq{final_dof_prior}. We use this prior 
extensively in Sections~\ref{section:sim} and~\ref{section:results} 
to model the Cepheid and SN populations,
although its utility is not limited to this particular problem.

\begin{figure*}
\includegraphics[width=2\figwidth]{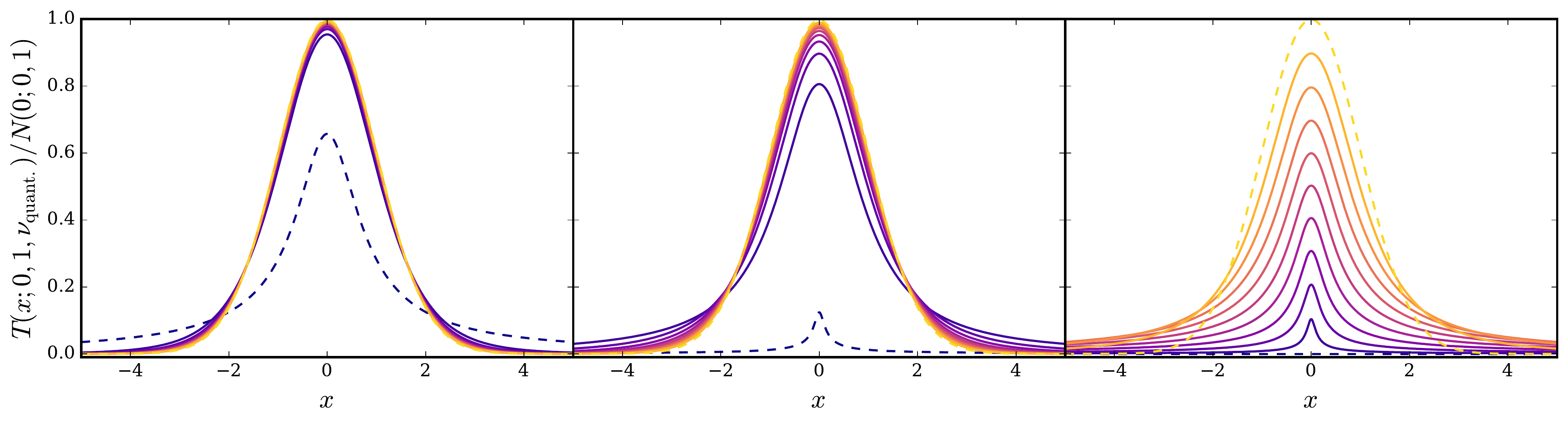}
\caption{Student-t distributions with degrees of freedom corresponding to the 0.1$^{\rm st}$, 10$^{\rm th}$, 20$^{\rm th}$, $\ldots$, 80$^{\rm th}$, 90$^{\rm th}$, and 99.9$^{\rm th}$ percentiles of the gamma (left), exponential (centre) and uniform-$\tpeak$ priors (right). Dark purple (light orange) quantiles have small (large) degrees of freedom; the extreme quantiles are indicated by dashed lines. The curves are normalized such that Gaussian distributions have unit amplitude.}
\label{figure:dof_prior_quantiles}
\end{figure*}


\bsp	
\label{lastpage}
\end{document}